\documentclass[aps,prb,superscriptaddress,floatfix,twocolumn]{revtex4}
\usepackage{amsfonts,amsmath,amssymb,bm}
\usepackage{graphicx,graphics,float}
\usepackage{color}
\usepackage[colorlinks,citecolor=blue,linkcolor=blue]{hyperref}
\newcommand{\be}{\begin{equation}}
\newcommand{\ee}{\end{equation}}
\newcommand{\bea}{\begin{eqnarray}}
\newcommand{\eea}{\end{eqnarray}}

\begin{document}

\title{Enhancing thermophotovoltaic performance using graphene-BN-InSb near-field heterostructures}

\author{Rongqian Wang}
\address{School of Physical Science and Technology \& Collaborative Innovation Center of Suzhou Nano Science and Technology, Soochow University, Suzhou 215006, China}

\author{Jincheng Lu}
\address{School of Physical Science and Technology \& Collaborative Innovation Center of Suzhou Nano Science and Technology, Soochow University, Suzhou 215006, China}

\author{Jian-Hua Jiang}\email{jianhuajiang@suda.edu.cn ,joejhjiang@hotmail.com}
\address{School of Physical Science and Technology \& Collaborative Innovation Center of Suzhou Nano Science and Technology, Soochow University, Suzhou 215006, China}
\address{Department of Physics, University of Toronto, 60 St. George Street, Toronto, Ontario, Canada M5S 1A7}

\date{\today}
\begin{abstract}
Graphene---hexagonal-boron-nitride---InSb near-field structures are designed and optimized to enhance the output power and energy
efficiency of the thermophotovoltaic systems working in the temperature range of common industrial waste heat, $400~\rm K \sim 800~\rm K$,
which is also the working temperature range for conventional thermoelectric devices. We show that the optimal output electric power can reach $3.5\times10^{4} \rm\ W/\rm m^2$ for the system with a graphene---hexagonal-boron-nitride heterostructure emitter and a graphene-covered
InSb cell, whereas the best efficiency is achieved by the system with the heterostructure emitter and an uncovered InSb cell (reaching to $27\%$
of the Carnot efficiency). These results show that the performances of near-field thermophotovoltaic systems can be comparable with or even
superior than the state-of-art thermoelectric devices. The underlying physics for the significant enhancement of the thermophotovoltaic
performance is understood as due to the resonant coupling between the emitter and the cell, where the surface plasmons in graphene and
surface phonon-polaritons in boron-nitride play important roles. Our study provides a stepping stone for future high-performance
thermophotovoltaic systems.
\end{abstract}

\maketitle

\section{Introduction}
Thermophotovoltaic (TPV) cells, as an emergent breed of clean energy resource, have attracted a wide range of research interest due to their potential high-performances~\cite{shockley1961detailed,swanson1980thermophotovoltaic,fraas1995spectral,wojtczuk1995x,hamlen1996thermophotovoltaic,coutts1998thermophotovoltaics,martin2004temperature,nagashima2007germanium,fraas2000three,sulima2001fabrication,wu2012metamaterial,chan2013toward,liao2016efficiently,Tervo2018}. In general, a TPV system consists of a thermal radiation emitter and an infrared photovoltaic cell which converts thermal radiations into electric power. The emitter can be heated either by the sun or by industrial waste heat, giving rise to different applications. Far-field thermal radiation is limited by the Stefan-Boltzmann law, results in reduced output power, particularly when the temperature of the emitter is below 1000~K. Most of the existing studies are focused on the regime where the emitter has very high temperature, e.g., $T_{\rm emit}>1000~\rm K$, which corresponds to solar radiations or heat radiations from a secondary thermal emitter which receives the solar radiation energy.

In this work, we focus on the situations where the temperature of the thermal emitter is in the range of common industrial
waste heat, $400~\rm K <\it T_{\rm emit}<\rm 800~\rm K$. At such a temperature range, thermal radiation is strongly limited
by the Stefan-Boltzmann law, leading to suppressed output power and energy efficiency for the TPV systems. One can
turn to the near-field effect~\cite{cravalho1967effect,xu1994Thermal,xu1994Proposal,xu1994Energy,xu1994heat,volokitin2001radiative,mulet2002enhanced,hu2008near}
to enhance thermal radiation flux, which may enable appealing energy efficiency and output power. However, the near-field
effect becomes ineffective if the tunneling through the vacuum gap is inefficient for photons that carry most of the
heat energy, i.e., photons of angular frequencies $\omega \sim k_{\rm B}T_{\rm emit}/\hbar$. One way to overcome this obstacle is
to exploit surface-polariton enhanced near-field radiations~\cite{ilic2012overcoming,messina2013graphene,Jinghua_graphene,svetovoy2012plasmon,svetovoy2014graphene}.
Graphene and hexagonal-boron-nitride (h-BN) provide
surface plasmon polaritons (SPPs) and surface phonon polaritons (SPhPs)~\cite{Dai_hBN,bnnt1,expara_hBN,bnnt2} due to strong light-matter
interactions~\cite{prx2014}. These surface polaritons are right in the
mid infrared frequency range that fits the peak thermal radiation frequencies of emitters in the temperature range of
$400~\rm K <\it T_{\rm emit}<\rm 800~\rm K$. The narrow band gap indium antimonide (InSb) $p$-$n$ junctions can be used as the
TPV cells to convert such mid infrared thermal radiation into electricity. However, there is still considerable
mismatch between the optical property of InSb and that of graphene or h-BN, which significantly reduces the near-field radiation.
It has been shown that a sheet of graphene covered on the InSb $p$-$n$ junction can strongly improve the coupling between the emitter and
the TPV cell and thus enhance the near-field thermal radiation~\cite{messina2013graphene}. Here, we propose to use hBN-graphene heterostructures~\cite{SPPPs1,SPPPs2,woessner2015retime,Bo_JHT,Bo_PRB,Sailing_ACS} as the emitter and
the graphene-covered InSb $p$-$n$ junction as the TPV cell. We find that such a design leads to significantly improved performance as compared
to the existing studies~\cite{messina2013graphene,heavens1991optical,knittl1976optics}. The main difference between our near-field TPV (NTPV)
systems and that in Ref.~[\onlinecite{messina2013graphene}] is that the thermal emitter in our system is made of h-BN---graphene heterostructures,
making our NTPV systems more powerful and efficient than the NTPV systems considered in Ref.~[\onlinecite{messina2013graphene}] where the
emitter is made of bulk h-BN.

In this paper, we examine the performances of four different NTPV configurations: (i) h-BN-InSb cell
(denoted as hBN-InSb, with the mono-structure bulk h-BN being the emitter and the uncovered InSb junction being the cell), (ii)
h-BN-graphene/InSb cell (denoted as hBN-G/InSb, with the bulk h-BN being the emitter and the graphene-covered InSb junction
as the cell), (iii) h-BN/graphene-InSb cell (denoted as fhBN/G-InSb, with the h-BN/graphene heterostructure film being the emitter and the
uncovered InSb junction as the cell), and (iv) h-BN/graphene-graphene/InSb (denoted as fhBN/G-G/InSb, with the h-BN/graphene
heterostructure film being the emitter and the graphene-covered InSb junction as the cell). We study and compare their performances for
various conditions to optimize the performance of the NTPV system and to reveal the underlying physical mechanisms toward
high-performance NTPV systems working at the temperature range of $400~{\rm K}<T_{\rm emit}<800~{\rm K}$. Although it has been
shown that a single graphene sheet can act as an excellent thermal emitter~\cite{ilic2012overcoming,lim2015graphene} which gives
high power density for the NTPV system. However, such a structure is technologically challenging and the induced energy efficiency
is considerably lower than the NTPV systems considered in this work, despite that the power density is comparable with our NTPV systems.

We remark that our findings are consistent with the study in Ref.~[\onlinecite{Jiang2018Near}] where the synergy
between near-field thermal radiation and inelastic thermoelectricity\cite{inTE1,inTE2,inTE3,pre2014,njp2013,inTE4} is shown
to have considerably improved performance even in the linear-transport regime. On the other hand, our work is also based on the previous
studies where the near-field effects are shown to improve the heat radiation flux by orders-of-magnitude by using infrared hyperbolic
metamaterials~\cite{SPPPs1,SPPPs2,woessner2015retime,Bo_JHT,Bo_PRB,Sailing_ACS}.

This paper is structured as follows. In Sec.~\ref{model and formula}, we describe our near-field TPV system, introduce the optical properties
of the emitter and the InSb cell. We clarify the radiative heat flux exchanged between the emitter and the cell. We also recall the basic formulations
of the photo-induced current, electric power and energy efficiency of the NTPV cell. In Sec.~\ref{results and discussion}, we
study the performances of the four different NTPV systems at various conditions to search for high-performance NTPV systems.
Finally, we summarize and conclude in Sec.~\ref{conlusions}.

\section{System and model} \label{model and formula}
\subsection{Near-field thermal radiation}\label{nfradiation}

\begin{figure}
\begin{center}
\includegraphics[width=3.2 in,height=1.8 in]{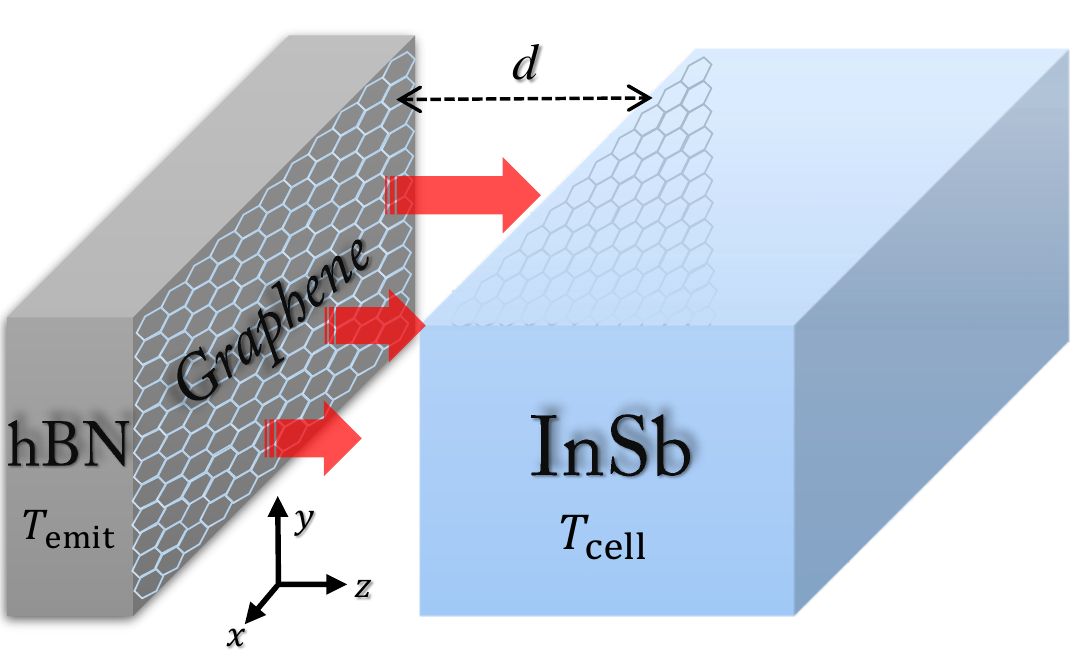}
\caption{(Color online)\ Schematic representation of the NTPV system. A thermal emitter of temperature $T_{\rm emit}$ made of h-BN/graphene heterostructure is placed in the proximity of a thermophotovoltaic cell of temperature $T_{\rm cell}$ made of InSb. The emitter-cell distance is kept at $d$. The red arrows represent the heat flux radiated from the emitter to the cell. The coordinate axes on the left side shows the in-plane (parallel to the $x-y$ plane) and out-of plane (perpendicular to the $x-y$ plane) directions.}\label{model}
\end{center}
\end{figure}

The proposed NTPV system is presented in Fig.~\ref{model}. The emitter is a graphene covered h-BN film of thickness $h$, kept at temperature
$T_{\rm emit}$. The TPV cell is made of an InSb $p$-$n$ junction, kept at temperature $T_{\rm cell}$, which is also
covered by a layer of graphene. The thermal radiation from the emitter is absorbed by the cell and converted
into electricity via infrared photoelectric conversion. In this system, plasmons in graphene can interact strongly with infrared photons
and give rise to SPPs. Meanwhile, phonons in h-BN can also couple strongly with photons
and leads to SPhPs~\cite{Dai_hBN,bnnt1,SPPPs1,SPPPs2,bnnt2,woessner2015retime,Jiang:17}. These emergent quasi-particles
due to strong light-matter interaction can dramatically enhance thermal radiation, particularly via the near-field effect~\cite{SPPPs1,SPPPs2,woessner2015retime,Bo_JHT,Bo_PRB,Sailing_ACS}.
Because of their coincident frequency ranges, the SPPs and SPhPs hybrid with each other when graphene is placed
together with h-BN, leading to hybrid polaritons called surface plasmon-phonon polaritons (SPPPs)~\cite{SPPPs1,SPPPs2}. SPPPs have been shown to be useful in improving near-field heat transfer
between two h-BN/graphene heterostructures~\cite{Bo_JHT,Bo_PRB,Sailing_ACS}. The radiative heat transfer
is optimized when the emitter and absorber are made of the same material, which leads to a resonant energy exchange between the emitter and the absorber~\cite{joulain2005surface}. In order to enable such resonant energy exchange in our NTPV cell, we add another layer of graphene onto the InSb cell, as shown in Fig.~\ref{model}.

The propagation of electromagnetic waves in these materials is described by the Maxwell equations, in which the dielectric functions of these materials
are the key factors that determine the propagation of the electromagnetic waves. The dielectric function of h-BN is described by a Drude-Lorentz
model, which is given by~\cite{SPPPs2}
\begin{equation}
\begin{aligned}
\varepsilon_m = \varepsilon_{\infty,m}\left(1+ \frac{\omega^2_{\rm LO,\it m}-\omega^2_{\rm TO,\it m}}{\omega^2_{\rm TO,\it m}-i\gamma_m\omega-\omega^2}\right),
\end{aligned}
\end{equation}
where $m = {\rm \parallel, \perp}$ denotes the out-of-plane and the in-plane directions, respectively. The in-(out-of-)plane direction is determined by
whether the electric field is perpendicular (parallel) to the optical axis of the h-BN film (h-BN is a uniaxial crystal, and the optical axis is in the $z$ direction,
as defined in Fig.~\ref{model}). $\varepsilon_{\infty,m}$ is the high-frequency relative permittivity, $\omega_{\rm TO}$ and $\omega_{\rm LO}$ are the transverse and
longitudinal optical phonon frequencies, respectively. $\gamma_m$ is the damping constant of the optical phonon modes. The values of these
parameters are chosen as those determined by experiments~\cite{geick1966normal,expara_hBN}, $\varepsilon_{\infty,\rm \perp} = 4.87$, $\omega_{\rm TO,\perp} = 1370$ $\rm cm^{-1}$,
$\omega_{\rm LO,\perp} = 1610$ $\rm cm^{-1}$ and $\gamma_{\perp} = 5$ $\rm cm^{-1}$ for in-plane phonon modes, $\varepsilon_{\infty,\rm \parallel}
= 2.95$, $\omega_{\rm TO,\parallel} = 780$ $\rm cm^{-1}$, $\omega_{\rm LO,\parallel} = 830$ $\rm cm^{-1}$ and $\gamma_{\parallel} = 4$ $\rm cm^{-1}$
for out-of-plane phonon modes. There are certain frequency ranges where the in-plane and out-of-plane dielectric functions of h-BN
have opposite signs, leading to the hyperbolicity of isofrequency contour, which makes h-BN a natural hyperbolic metamaterial~\cite{SPPPs2}.

The optical property of graphene sheet is described here by a modeled conductivity, $\sigma_{\rm g}$, which sums over both the intraband $\sigma_{\rm D}$ and interband $\sigma_{\rm I}$ contributions, i.e., $\sigma_{\rm g} = \sigma_{\rm D}+\sigma_{\rm I}$, given respectively by~\cite{conductivity_graphene}
\begin{equation}
\begin{aligned}
\sigma_{\rm D} = \frac{i}{\omega+\frac{i}{\tau}}\frac{2e^2k_{\rm B}T}{\rm \pi\hbar^2}\ln\left[2\cosh\left(\frac{\mu_{\rm g}}{2k_{\rm B}T}\right)\right],
\end{aligned}
\end{equation}
and
\begin{equation}
\begin{aligned}
\sigma_{\rm I} = \frac{e^2}{4\rm \hbar}[G\left(\frac{\rm \hbar\omega}{2}\right)+i\frac{4\hbar\it \omega}{\pi}\int_{0}^{\infty}\frac{G(\xi)-G(\frac{\rm \hbar\it \omega}{2})}{(\rm \hbar\it \omega)^2-\rm{4}\it\xi^2}d\xi],
\end{aligned}
\end{equation}
where $G(\xi)=\frac{\sinh\left(\frac{\xi}{k_{\rm B}T}\right)}{\cosh\left(\frac{\xi}{k_{\rm B}T}\right)+\cosh\left(\frac{\mu_{\rm g}}{k_{\rm B}T}\right)}$ is a dimensionless function. $\mu_{\rm g}$ is the chemical potential of the graphene sheet, which can be readily tuned via a gate-voltage~\cite{yan2013damping}. $\tau$ is the electron scattering time and chosen as 100 fs in our calculations~\cite{woessner2015retime}. $T$ is the temperature of the graphene sheet. $e$, $k_{\rm B}$ and $\hbar$ are the electron charge, Boltzmann constant and the reduced Planck constant, separately.

The dielectric function of the InSb cell is given by~\cite{messina2013graphene}
\begin{equation}
\begin{aligned}
\varepsilon = \left(n+\frac{i\rm c \it\alpha(\omega)}{2\omega}\right)^2,
\end{aligned}
\end{equation}
where $n=4.12$ is the refractive index and $c$ is the speed of light in vacuum. $\alpha(\omega)$ is a step-like function which describes the absorption probability of the incident photons, and it is given by $\alpha(\omega)=0$ for $\omega < \omega_{\rm gap}$ and $\alpha(\omega)=\alpha_0\sqrt{\frac{\omega}{\omega_{\rm gap}}-1}$ for $\omega > \omega_{\rm gap}$ with $\alpha_0$ = 0.7 $\mu\rm m^{-1}$ being the absorption coefficient and $\omega_{\rm gap}=\frac{E_{\rm gap}}{\hbar}$ being the angular frequency corresponding to the band gap of InSb. The gap energy $E_{\rm gap}$ of InSb is temperature dependent, which is given by $E_{\rm gap}=0.24-6\times10^{-4}\frac{T^2}{T+500}$~\cite{shur1996handbook}. For instance, for $T_{\rm cell}=320\ \rm K$, the gap energy $E_{\rm gap}$ is nearly 0.17 eV and the corresponding frequency is about $2.5\times10^{14}$ Hz.

The near-field radiative heat flux exchanged between the emitter and the cell is given by ~\cite{polder1971theory,pendry1999radiative},
\begin{equation}
\begin{aligned}
P_{\rm rad}=\frac{\delta\Theta(T_{\rm emit},T_{\rm cell},\omega,\Delta\mu)}{4\pi^2}\sum_{j}\int kdk \zeta_j\left(\omega,k\right),\label{Planck oscilator}
\end{aligned}
\end{equation}
where $\delta\Theta\left(T_{\rm emit},T_{\rm cell},\omega,\Delta\mu\right)=\Theta_1\left(T_{\rm emit},\omega\right)-\Theta_2\left(T_{\rm cell},\omega,\Delta\mu\right)$, with $\Theta_1\left(T_{\rm emit},\omega\right)=\frac{\hbar\omega}{\exp{\left(\frac{\hbar\omega}{k_{\rm B}T_{\rm emit}}\right)}-1}$ and $\Theta_2\left(T_{\rm cell},\omega,\Delta\mu\right)=\frac{\hbar\omega}{\exp{\left(\frac{\hbar\omega-\Delta\mu}{k_{\rm B}T_{\rm cell}}\right)}-1}$ being the Planck mean oscillator energy of blackbody at temperature $T_{\rm emit}$ and $T_{\rm cell}$, respectively. $\Delta\mu$ is the electrochemical potential difference across the $p$-$n$ junction in the TPV cell. $k$ denotes the magnitude of the in-plane wavevector of thermal radiation waves and $\zeta_j(\omega,k)$ is called the photon transmission coefficient for the $j$-polarization $(j=s,p)$, which represents the probability for a photon with angular frequency $\omega$ and wavevector $k$ to tunnel through the vacuum gap from the emitter to the cell. The transmission coefficient for each polarization sums over the contribution from both propagating and evanescent waves~\cite{mulet2002enhanced},
\begin{equation}
\zeta_j(\omega,k)=
\begin{cases}
\frac{\left(1-\left|r_{\rm emit}\right|^2\right)\left(1-\left|r_{\rm cell}\right|^2\right)}{\left|1-r_{\rm emit}^j r_{\rm cell}^j \exp(2ik_zd)\right|}, & k<\frac{\omega}{c} \\

             \frac{4{\rm Im}\left(r_{\rm emit}^j\right){\rm Im}\left(r_{\rm cell}^j\right)\exp(2ik_zd)}{\left|1-r_{\rm emit}^j r_{\rm cell}^j \exp(2ik_zd)\right|}, & k>\frac{\omega}{c} \label{photon tunneling}
\end{cases}
\end{equation}
where $k_z=\sqrt{\frac{\omega^2}{c^2}-k^2}$ is the normal component of the wavevector in vacuum. $r_{\rm emit}^j$ and $r_{\rm cell}^j$ ($j=s,p$) are
the reflection coefficients of the emitter and the cell, respectively, which measure the amplitude ratio between the reflected and incident waves at the
interface of the emitter and the cell, respectively. These coefficients can be calculated by the scattering matrix approach where the optical
interference in the heterostructure film is also taken into account~\cite{Zhang2007Nano},
\begin{equation}
\begin{aligned}
r_{\rm emit}^s=\frac{r_{12}^s+\left(1+r_{12}^s+r_{21}^s\right)r_{23}^s\exp(2ik_{z,2}^sh)}{1-r_{21}^sr_{23}^s\exp(2ik_{k,2}^sh)},
\end{aligned}
\end{equation}

\begin{equation}
\begin{aligned}
r_{\rm cell}^s = \frac{k_{z}^s-k_{z,\rm cell}^s-\mu_0\sigma_{\rm g}\omega}{k_{z}^s+k_{z,\rm cell}^s+\mu_0\sigma_{\rm g}\omega}
\end{aligned}
\end{equation}
for s-polarization.

And
\begin{equation}
\begin{aligned}
r_{\rm emit}^p=\frac{r_{12}^p+\left(1-r_{12}^p-r_{21}^p\right)r_{23}^p\exp(2ik_{z,2}^ph)}{1-r_{21}^pr_{23}^p\exp(2ik_{k,2}^ph)},
\end{aligned}
\end{equation}

\begin{equation}
\begin{aligned}
r_{\rm cell}^p=\frac{\varepsilon_{\rm cell}^{\perp}k_{z}^p-k_{z,\rm cell}^p+\frac{\sigma_{\rm g}k_{z}^pk_{z,\rm cell}^p}{\varepsilon_{\rm 0}\omega}}{\varepsilon_{\rm cell}^{\perp}k_{z}^p+k_{z,\rm cell}^p+\frac{\sigma_{\rm g}k_{z}^pk_{z,\rm cell}^p}{\varepsilon_{\rm 0}\omega}}
\end{aligned}
\end{equation}
for p-polarization. Where 1, 2, and 3 are the indexes for the vacuum region above h-BN film, the h-BN film region, and the vacuum region below h-BN film,respectively, as defined in Ref.~[\onlinecite{Bo_JHT}]. $r_{ab}^{s,p}$ is the reflection coefficient between media $a$ and $b$ ($a,b=1,2,3$) for either $s$ or $p$ polarization, which is given by
\begin{equation}
\begin{aligned}
& r_{ab}^s=\frac{k_{z,a}^s-k_{z,b}-\sigma_{\rm g}\omega\mu_0}{k_{z,a}^s+k_{z,b}+\sigma_{\rm g}\omega\mu_0},\\
& r_{ab}^p=\frac{k_{z,a}^p\varepsilon_b^{\perp}-k_{z,b}\varepsilon_a^{\perp}+k_{z,a}^p k_{z,b}\frac{\sigma_{\rm g}}{\omega\varepsilon_0}}{k_{z,a}^p\varepsilon_b^{\perp}+k_{z,b}\varepsilon_a^{\perp}+k_{z,a}^p k_{z,b}\frac{\sigma_{\rm g}}{\omega\varepsilon_0}},
\end{aligned}
\end{equation}
when there is a graphene layer in between media $a$ and $b$,
\begin{equation}
\begin{aligned}
& r_{ab}^s=\frac{k_{z,a}^s-k_{z,b}}{k_{z,a}^s+k_{z,b}},\\
& r_{ab}^p=\frac{k_{z,a}^p\varepsilon_b^{\perp}-k_{z,b}\varepsilon_a^{\perp}}{k_{z,a}^p\varepsilon_b^{\perp}+k_{z,b}\varepsilon_a^{\perp}},
\end{aligned}
\end{equation}
when there is no graphene in between media $a$ and $b$. Here $k_{z,i}^s=\sqrt{\varepsilon_i^{\perp}\frac{\omega^2}{c^2}-k^2}$
and $k_{z,i}^p=\sqrt{\varepsilon_i^{\perp}\frac{\omega^2}{c^2}-\frac{\varepsilon_i^{\perp}}{\varepsilon_i^{\parallel}}k^2} \left(i=a, b, \rm cell\right)$ are the $z$-component wavevectors for media $i$ for $s$ and $p$ polarization, respectively. $\varepsilon_{i}^{\perp}$ and $\varepsilon_{i}^{\parallel}  \left(i=a, b, \rm cell\right)$ are the in-plane and out-of plane components of the relative dielectric tensor. For isotropic medium like vacuum and InSb, $\varepsilon_1^{\perp}=\varepsilon_1^{\parallel}=\varepsilon_3^{\perp}=\varepsilon_3^{\parallel}=1$, and $\varepsilon_{\rm cell}^{\perp}=\varepsilon_{\rm cell}^{\parallel}$, where the dielectric function of the InSb cell is defined before. $\varepsilon_0$ and $\mu_0$ are the permittivity and permeability for vacuum, respectively.

\subsection{Characterizations of TPV cells} \label{defperformance}
 When the TPV cell is located at a distance $d$ which is on the order of or smaller than the thermal wavelength $\lambda_{\rm th}=2\pi \hbar c/k_{\rm B}T_{\rm emit}$ from the emitter, the radiative energy transfer will be enhanced due to the presence of evanescent waves~\cite{ilic2012overcoming}, which exists only in the vicinity of the emitter, because of the exponential decay from the interface. The enhanced radiation with energy greater than the band gap $E_{\rm gap}$ will be absorbed by the TPV cell, creating electron-hole pairs and leading to electric current flow and output power generation. The induced electric current density can be written as~\cite{shockley1961detailed,ashcroft2010IVcurve}
\begin{equation}
\begin{aligned}
I_{\rm e}=I_{\rm ph}-I_0[\exp(V/V_{\rm cell})-1], \label{ecurrent}
\end{aligned}
\end{equation}
where $V=\Delta\mu/e$ is the voltage bias across the TPV cell, $V_{\rm cell}=\frac{k_{\rm B}T_{\rm cell}}{e}$ is a voltage which measures the temperature of the cell~\cite{shockley1961detailed}. $I_{\rm ph}$ and $I_0$ are called the photo-induced current density and reverse saturation current density, respectively. The reverse saturation current density (also named the dark current density, the electric current without light) is determined by the diffusion of minority carriers in the InSb $p$-$n$ junction, which is given by
\begin{equation}
\begin{aligned}
I_0=en_{\rm i}^2\left(\frac{1}{N_{\rm A}}\sqrt{\frac{D_{\rm e}}{\tau_{\rm e}}}+\frac{1}{N_{\rm D}}\sqrt{\frac{D_{\rm h}}{\tau_{\rm h}}} \right),\label{rcurrent}
\end{aligned}
\end{equation}
 where $n_{\rm i}=\sqrt{N_{\rm c}N_{\rm v}}\exp(-\frac{E_{\rm gap}}{2k_{\rm B}T_{\rm cell}})$ $(\rm cm^{-3})$ is the intrinsic carrier concentration, which is dependent on the temperature of the TPV cell~\cite{shur1996handbook}. $N_{\rm c}$ and $N_{\rm v}$ are temperature-dependent effective density of states in the conduction and valance bands, respectively given by $N_{\rm c}=8\times10^{12}\times T_{\rm cell}^{\frac{3}{2}}$ $(\rm cm^{-3})$ and $N_{\rm v}=1.4\times10^{15}\times T_{\rm cell}^{\frac{3}{2}}$ $(\rm cm^{-3})$. $N_{\rm A}$ and $N_{\rm D}$ are the $p$-region and $n$-region impurity concentrations, respectively. Numerical values are set as $N_{\rm A}=N_{\rm D}=10^{19}$ $(\rm cm^{-3})$. $D_{\rm e}$ and $D_{\rm h}$ are the diffusion coefficients of the electrons and holes, taken as $D_{\rm e}=186$ $\rm cm^2/s$ and $D_{\rm h}=5.21$ $\rm cm^2/s$\cite{lim2015graphene}. $\tau_{\rm e}$ and $\tau_{h}$ are the relaxation times of the electron-hole pairs in the $n$-region and $p$-region, correspondingly. Values for relaxation times are calculated by $\tau_{\rm e}=\tau_{h}=\frac{1}{\left(5\times10^{-26}\,{\rm cm^{-6}s^{-1}}\right)n_{\rm i}^2}$\cite{shur1996handbook}.

Photo-induced current is the electric current induced by the motion of photo-carriers. The photo-induced current density $I_{\rm ph}$ is given by~\cite{laroche2006JAP}
\begin{equation}
\begin{aligned}
I_{\rm ph}=e\int_{\omega_{\rm gap}}^{\infty}\frac{P_{\rm rad}\left(T_{\rm emit},T_{\rm cell},\omega,\Delta\mu \right)}{\hbar\omega}d\omega, \label{Iph}
\end{aligned}
\end{equation}
where the radiative spectral heat flux is defined in Sec. \ref{nfradiation}. Here we assume that all incident photons are absorbed, and each photon with an energy greater than the band gap creates one electron-hole pair, i.e., we assume $100\%$ quantum efficiency~\cite{messina2013graphene}. Under this assumption, we only consider the energy efficiency, which measures how much incident radiative heat can be transferred into electricity.
We keep this assumption which is widely adopted in the literature.

The output electric power density $P_{\rm e}$ is defined as the product of the net electric current density and the voltage bias,
\begin{equation}
\begin{aligned}
P_{\rm e}= -I_{\rm e}V, \label{epower}
\end{aligned}
\end{equation}
and the energy efficiency $\eta$ is given by the ratio between the output electric power density $P_{\rm e}$ and incident radiative heat flux $Q_{\rm inc}$,
\begin{equation}
\begin{aligned}
\eta= \frac{P_{\rm e}}{Q_{\rm inc}},\label{effi}
\end{aligned}
\end{equation}
where the incident radiative heat flux is given by
\begin{align}
& Q_{\rm inc}=\int_{0}^{\infty}\frac{d\omega}{4\pi^2}\Theta_{1}\left(T_{\rm emit},\omega \right)\sum_{j}\int kdk \zeta_j(\omega,k) \nonumber \\
& \hspace{1cm} -\int_{\omega_{\rm gap}}^{\infty}\frac{d\omega}{4\pi^2}\Theta_{2}\left(T_{\rm cell},\omega,\Delta\mu\right)\sum_{j}\int kdk \zeta_j(\omega,k) \nonumber \\
& \hspace{1cm} -\int_{0}^{\omega_{\rm gap}}\frac{d\omega}{4\pi^2}\Theta_{1}\left(T_{\rm cell},\omega\right)\sum_{j}\int kdk \zeta_j(\omega,k), \label{Pinc}
\end{align}

\begin{figure}
\centering\includegraphics[width=3.3 in,height=1.4 in]{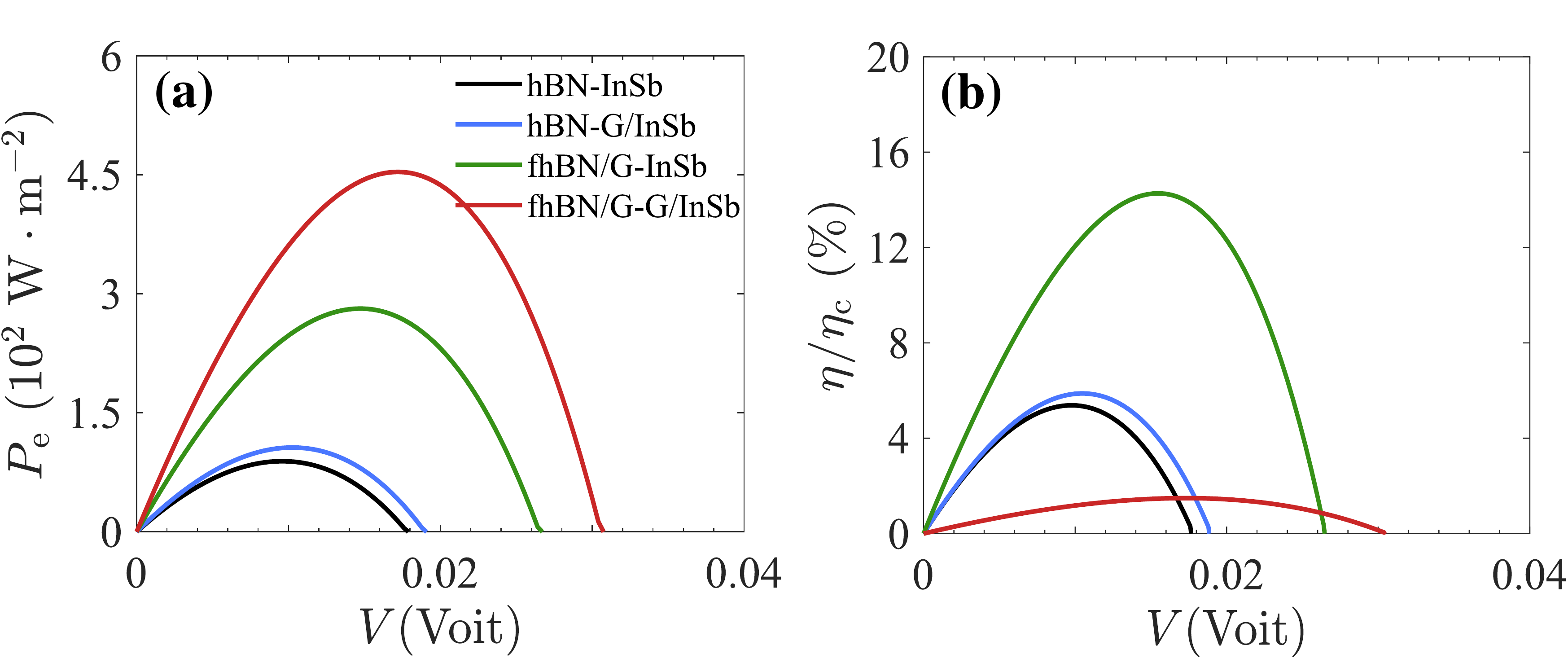}
\centering\includegraphics[width=3.3 in,height=1.4 in]{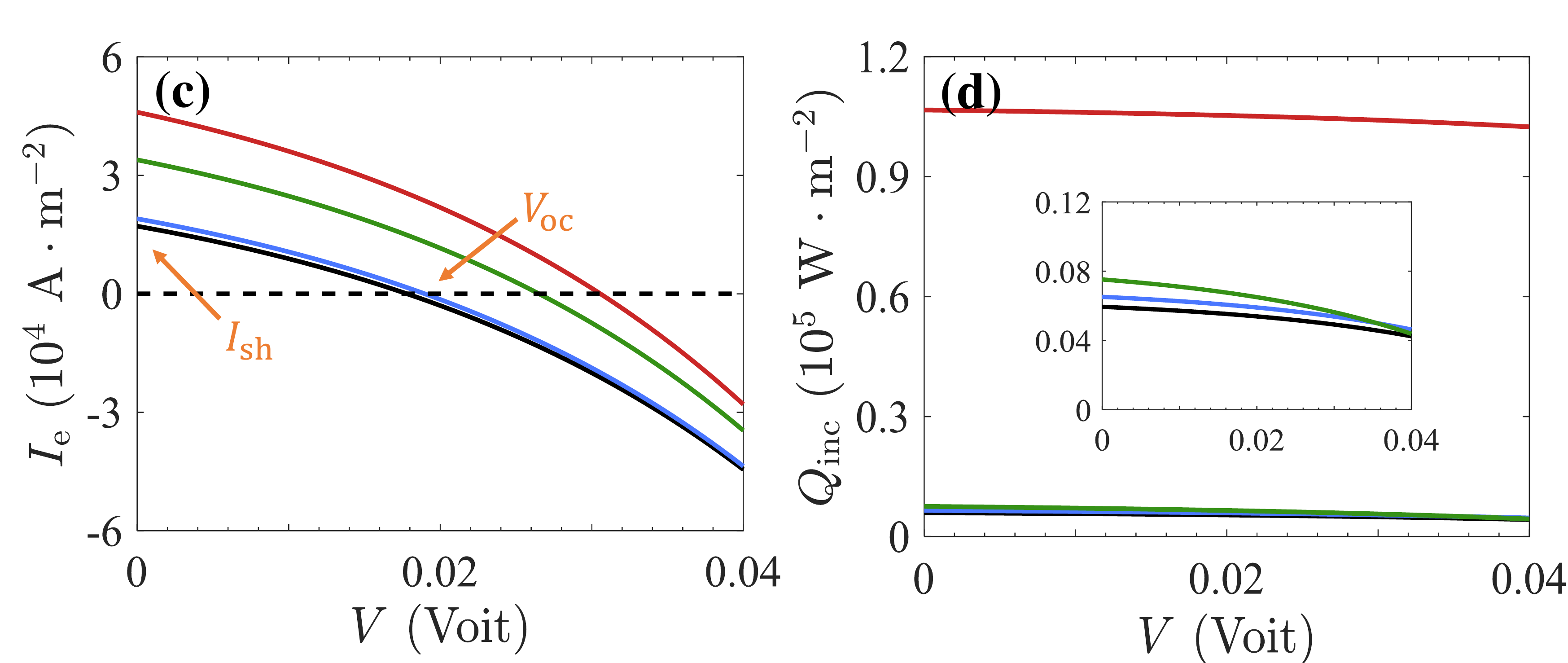}
\caption{(Color online)\ Performances of four NTPV cells. (a) Electric power density $P_{\rm e}$, (b) energy efficiency in units of Carnot efficiency ($\eta / \eta_{\rm c}$), (c) electric current density $I_{\rm e}$ and (d) incident radiative heat flux $Q_{\rm inc}$. The temperatures of the emitter and the cell are kept at $T_{\rm emit}=450$ K and $T_{\rm cell}=320$ K, respectively. The vacuum gap is $d=22$ nm, the thickness of h-BN is set as $h_{\rm bulk}=10000$ nm for bulk one, and $h_{\rm film}=20$ nm for h-BN thin film. The chemical potential of graphene is $\mu_{\rm g}=0.37$ eV. All parameters are same for the four configurations. The Carnot efficiency is given by $\eta_{\rm c}=1-\frac{T_{\rm cell}}{T_{\rm emit}}$.}\label{fig:performance}
\end{figure}

\section{Results and Discussions}\label{results and discussion}
\subsection{Performance of the designed NTPV systems}\label{nfperformance}
We first examine the electric power and the normalized energy efficiency in units of the Carnot efficiency ($\eta_{\rm c}$) for our proposed
near-field TPV cells. \ Fig.~\ref{fig:performance} shows the performances for the four different configurations, respectively denoted as hBN-InSb,
hBN-G/InSb, fhBN/G-InSb, and fhBN/G-G/InSb. The electric power density and efficiency are calculated through Eq.~(\ref{epower}) and (\ref{effi}),
respectively. It is noted that by adding a single layer of graphene either on the emitter or the TPV cell, the output electric power density is considerably
enhanced. Specifically, the enhancement factors of maximum electric power for fhBN/G-InSb cell (blue solid line) and hBN-G/InSb cell (green solid line)
are respectively given by $P_{\rm max}^{\rm hBN-G/InSb}=3.2P_{\rm max}^{\rm hBN-InSb}$ and $P_{\rm max}^{\rm fhBN/G-InSb}=
1.2P_{\rm max}^{\rm hBN-InSb}$. This enhancement induced by graphene (due to graphene SPPs) has been found in
Ref.~[\onlinecite{messina2013graphene}]. Notably, when the emitter and cell are both covered by graphene, the maximum electric power density
is significantly enhanced, with the enhancement factor $P_{\rm max}^{\rm fhBN/G-G/InSb}=5.1P_{\rm max}^{\rm hBN-InSb}$.

\begin{figure*}
\centering\includegraphics[width=5 in,height=2.3 in]{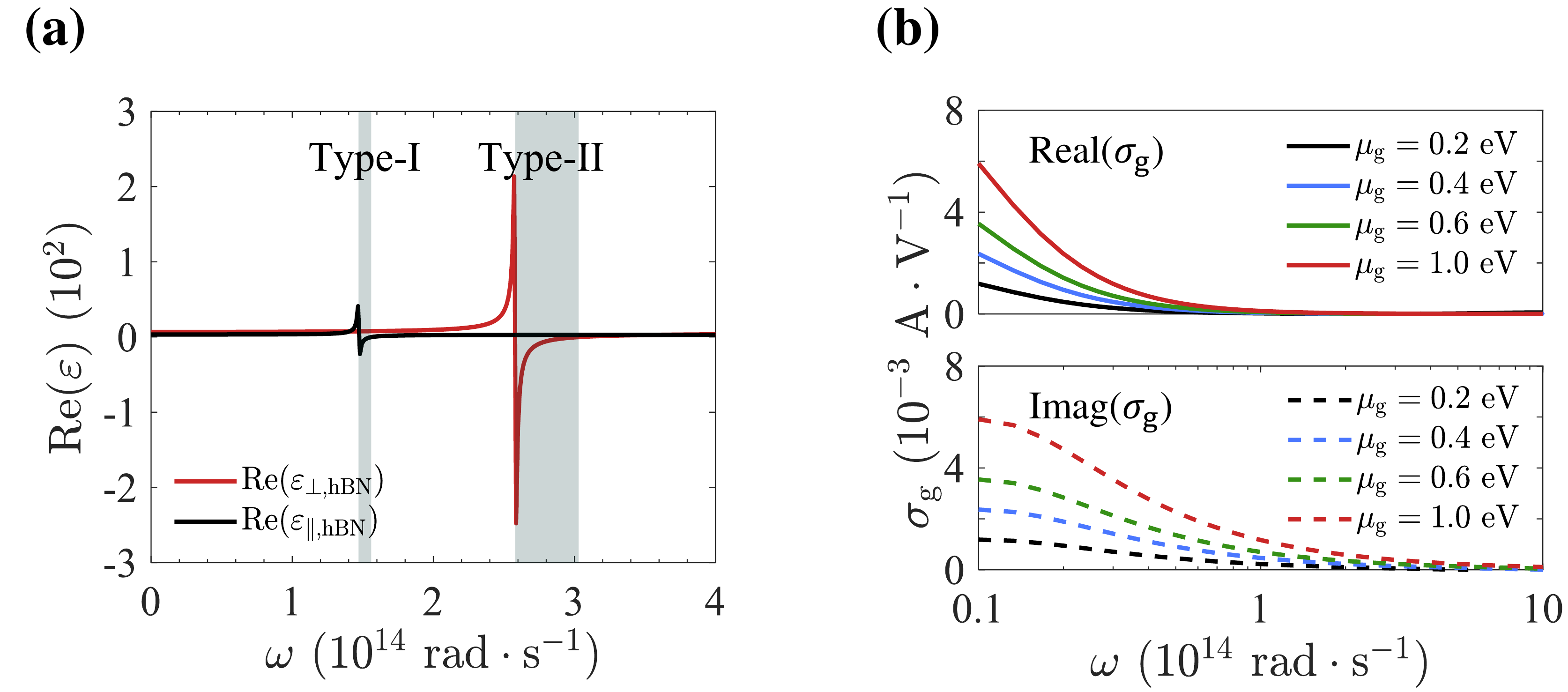}
\centering\includegraphics[width=5 in,height=2.3 in]{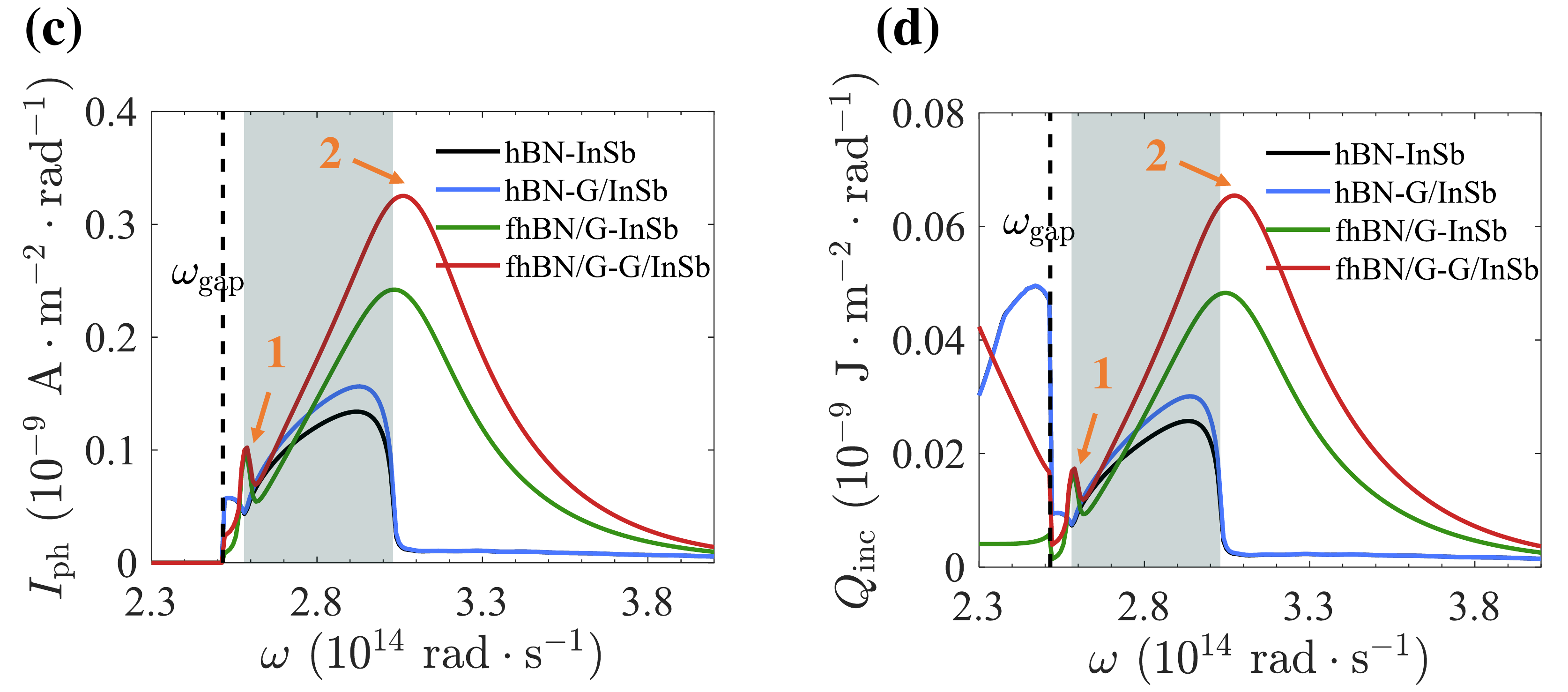}
\caption{(Color online)\ (a) Dielectric function $\varepsilon$ of h-BN, (b) conductivity of graphene $\sigma_{\rm g}$, (c) photo-induced current spectrums $I_{\rm ph}\left(\omega\right)$ and (d) incident radiative heat spectrums $Q_{\rm inc}\left(\omega\right)$ for the four configurations. The parameters are same with those in Fig.2. The two shaded areas represent type-I and type-II hyperbolic regions of h-BN,
respectively.}\label{fig:spectrum}
\end{figure*}

\begin{figure*}
\centering\includegraphics[width=5 in,height=2.3 in]{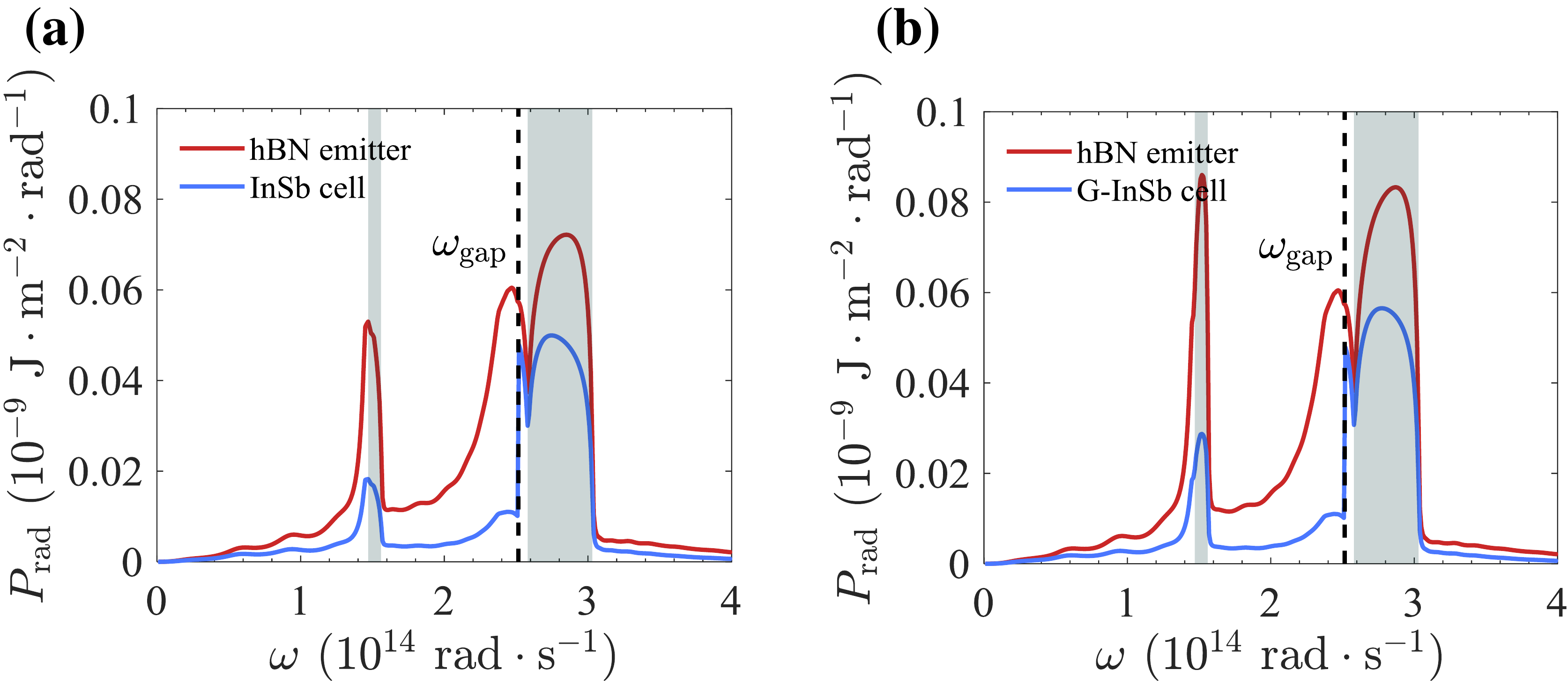}
\centering\includegraphics[width=5 in,height=2.3 in]{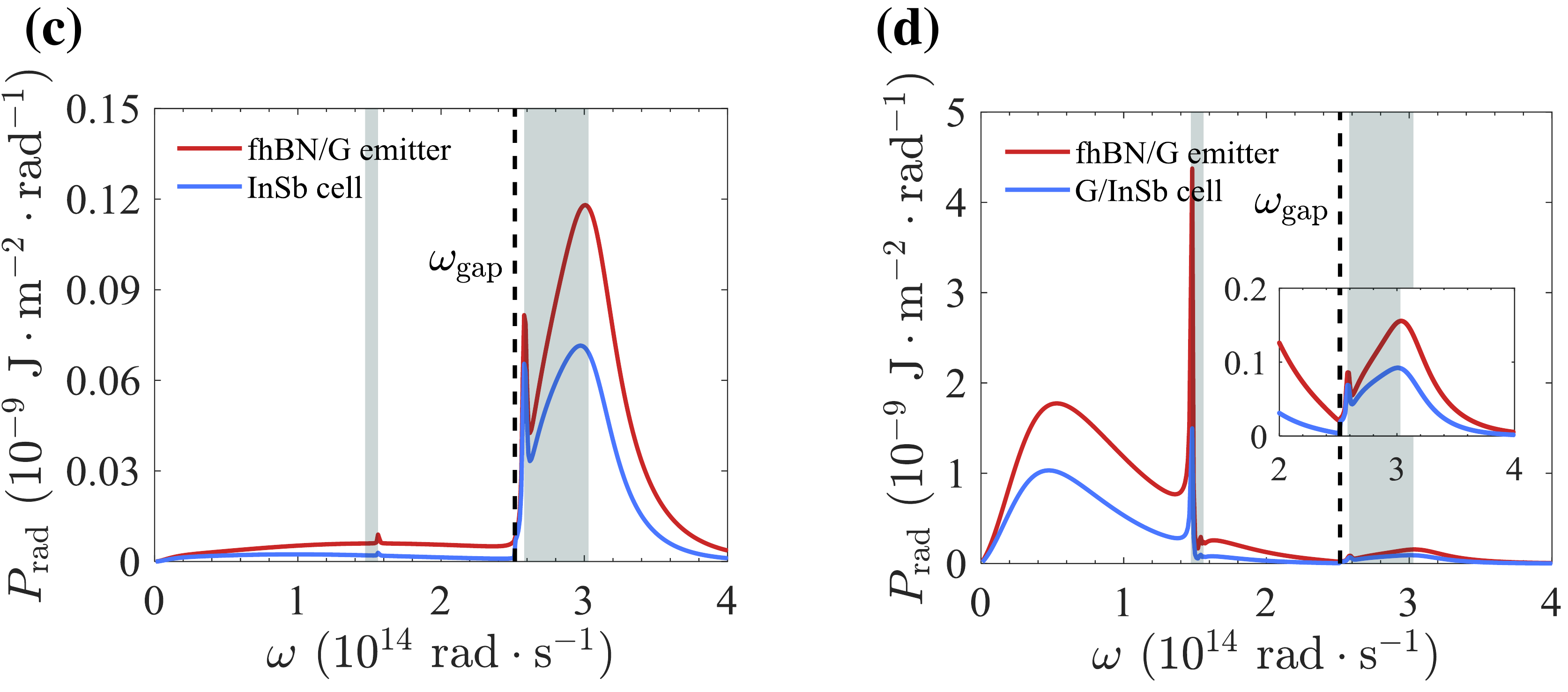}
\caption{(Color online)\ Radiative emission spectrums $P_{\rm rad}\left(\omega\right)$ for (a) hBN-InSb cell, (b) hBN-G/InSb cell, (c) fhBN/G-InSb cell and (d) fhBN/G-G/InSb cell. The parameters are same with those in Fig.2.}\label{fig:emission}
\end{figure*}

As indicated in Fig.~\ref{fig:performance}(b), the energy efficiency also shows strong enhancement for hBN-G/InSb and fhBN/G-InSb cells, with the
enhancement of the maximum efficiency $\eta_{\rm max}^{\rm hBN-G/InSb}=1.1\eta_{\rm max}^{\rm hBN-InSb}$ and
$\eta_{\rm max}^{\rm fhBN/G-InSb}=2.7\eta_{\rm max}^{\rm hBN-InSb}$, respectively. While for fhBN/G-G/InSb cell, contrary to the considerable
enhancement in output electric power, the energy efficiency for fhBN/G-G/InSb cell is substantially reduced.

As shown in Fig.~\ref{fig:performance}(c), the enhancement factors of short-circuit current densities
$I_{\rm sh}$ (when the voltage bias is zero) for hBN-G/InSb, fhBN/G-InSb and fhBN/G-G/InSb cells are about
$I_{\rm sh}^{\rm hBN-G/InSb}=1.1I_{\rm sh}^{\rm hBN-InSb}$ , $I_{\rm sh}^{\rm fhBN/G-InSb}=2.0I_{\rm sh}^{\rm hBN-InSb}$ and
$I_{\rm sh}^{\rm fhBN/G-G/InSb}=2.8I_{\rm sh}^{\rm hBN-InSb}$, respectively. Besides, the enhancement factors of open-circuit voltage
$V_{\rm oc}$ (when the electric current is zero) for the three covered cells are given by
$V_{\rm oc}^{\rm hBN-G/InSb}=1.1V_{\rm oc}^{\rm hBN-InSb}$ , $V_{\rm oc}^{\rm fhBN/G-InSb}=1.6V_{\rm oc}^{\rm hBN-InSb}$ and
$V_{\rm oc}^{\rm fhBN/G-G/InSb}=1.8V_{\rm oc}^{\rm hBN-InSb}$, respectively. The product of the short-circuit current $I_{\rm sh}$ and
the open-circuit voltage gives a good estimation of the output power. Therefore, the enhancements of the short-circuit current and
the open-circuit voltage lead to the significant improvement of the output power density shown in Fig.~\ref{fig:performance}(a).

The different manifestation of energy efficiency for graphene-covered near-field systems can be interpreted by the competition between the electric
power output and the radiative heat consumption. As shown in Fig.~\ref{fig:performance}(d), the incident radiative heat flux (given by Eq.~\eqref{Pinc})
is increased $Q_{\rm inc}^{\rm hBN-G/InSb}=1.1Q_{\rm inc}^{\rm hBN-InSb}$, and $Q_{\rm inc}^{\rm fhBN/G-InSb}=1.3Q_{\rm inc}^{\rm hBN-InSb}$
for the hBN-G/InSb cell and fhBN/G-InSb cell, respectively. The enhancement of the output electric power exceeds the increase of the radiative
heat consumption, leading to the improvement of energy efficiency. While for the fhBN/G-G/InSb cell, the radiative
heat consumption is significantly increased $Q_{\rm inc}^{\rm fhBN/G-G/InSb}=18Q_{\rm inc}^{\rm hBN-InSb}$, much higher than the improvement
of the electric power output, leading to suppressed energy efficiency.

To figure out the microscopic mechanisms for the enhancement of the output power, we study the spectral distributions of the photo-induced current $I_{\rm ph}\left(\omega\right)$ with the help of the dielectric function of h-BN and the optical conductivity of graphene. As shown in
Fig.~\ref{fig:spectrum}(a), two hyperbolic regions (labeled with type-I and
type-II) are marked with gray shades. For type-I hyperbolicity,
the corresponding frequency range is from $1.5\times10^{14} \rm\ Hz$ (corresponding to the frequency of transverse phonon vibrations,
$\omega_{\rm TO,\perp} = 780$ $\rm cm^{-1}$) to $1.6\times10^{14} \rm\ Hz$ (corresponding to the frequency of longitudinal phonon
vibrations, $\omega_{\rm LO,\perp} = 830$ $\rm cm^{-1}$). For type-II hyperbolicity, the corresponding
frequency range is from $\omega_{\rm TO,\perp} = 1370$ $\rm cm^{-1}$) to $3.1\times10^{14} \rm\ Hz$ (corresponding to the frequency of
longitudinal phonon vibrations, $\omega_{\rm LO,\perp} = 1610$ $\rm cm^{-1}$). The sharp peaks shown in these two regions indicate the
existence of SPhPs~\cite{jacob2014nanophotonics,geick1966normal}. While in Fig.~\ref{fig:spectrum}(b), the optical conductivities of graphene
at various chemical potentials show the broadband nature of the graphene SPPs~\cite{yin2016near}.

The photo-induced current spectrum $I_{\rm ph}\left(\omega\right)$ is given by Eq.~\eqref{Iph} but integrated over $k$ only. As shown in Fig.~\ref{fig:spectrum}(c), for hBN-InSb cell (black solid curve) the photo-induced current spectrum with a broad resonant peak is shown in the type-II hyperbolic regime (marked by the gray shade), which comes from the SPhPs. In hBN-G/InSb cell, the
graphene sheet on InSb cell induces a resonant absorption between the emitter and the cell, depicted by the higher peak (blue solid curve), with
about 20\% increase compared with the hBN-InSb cell. Since near-field radiation is dominated by evanescent wave couplings between the
emitter and the absorber, the resonant near-field absorption peak can be understood as due to the resonant coupling between the h-BN emitter
and the InSb absorber. The resonant near-field absorption peak gives the major contribution to the photo-induced current and the output electric
power.

Interestingly, there are two resonant peaks in the photo-induced current spectrum of the fhBN/G-InSb cell,
separately labeled as 1 and 2 in the Fig.~\ref{fig:spectrum}(c). And the corresponding frequencies are given by $\omega_1=2.6\times10^{14} \rm\ Hz$
and $\omega_2=3.1\times10^{14} \rm\ Hz$, respectively. The low-frequency resonance and the broad high-absorption band in the
type-II hyperbolic regime, are due to the hybrid polaritons named as the hyperbolic plasmon-phonon
polaritons (HPPPs), which are resulted from the coupling between the SPPs in graphene and the SPhPs in h-BN in the
type-II hyperbolic region~\cite{Dai_hBN}. The high-frequency absorption resonance outside the type-II region
is due to the surface plasmon-phonon polaritons (SPPPs), which come from the strong coupling between the broadband graphene plasmons
and the optical phonons outside the hyperbolic regions (i.e., regions without SPhPs)~\cite{SPPPs2,hajian2017hybrid}. As a consequence, the photo-induced current is much larger and the bandwidth
is much broader, which give rise to the major contribution in the photo-induced carrier generation and the electric power output.
Further enhancement of the photo-induced carrier generation can be realized by adding a monolayer graphene to the InSb cell, as shown
in Fig.~\ref{fig:performance}(c).

The incident radiative heat flux $Q_{\rm inc}\left(\omega\right)$ is examined in Fig.~\ref{fig:spectrum}(d). The overall trends of
$Q_{\rm inc}\left(\omega\right)$ are the same with the photo-induced current spectrums shown in Fig.~\ref{fig:spectrum}(c). However,
the incident radiative heat flux also includes the contributions from the below-band-gap near field heat transfer, which is quite different
for the four configurations.

In order to understand the below-band-gap near field heat transfer, we present the near-field heat flux in a very large frequency range
for both the emitter and the absorber in Fig.~\ref{fig:emission}. The difference between the heat fluxes flowing out of the emitter and the
absorber gives the radiative heat consumption, which consists of the useful part (above the InSb band gap) and the wasted part (below the
InSb band gap). The band gap frequency $\omega_{\rm gap}$ of the InSb cell is marked by the black dashed line.

Comparing between Fig.~\ref{fig:emission}(a) and Fig.~\ref{fig:emission}(b), one can see that adding a graphene layer to the InSb cell increases
both the useful and the wasted parts of the radiative heat flux, leading to slightly improved output power and energy efficiency.
The best energy efficiency improvement comes from the configuration with a graphene layer added to the h-BN emitter. As shown in
Fig.~\ref{fig:emission}(c), for this configuration the useful part of the radiative heat flux is significantly increased, while the wasted heat
flux is reduced. Therefore, the fhBN/G-InSb cell yields a significantly improved energy efficiency and output power. For the configuration
with fhBN/G emitter and graphene-covered InSb junction, as shown in Fig.~\ref{fig:emission}(d), the wasted radiative heat flux is dramatically
increased. Although this configuration has the largest output power, the energy efficiency is the lowest among those four configurations.

\begin{figure}
\centering\includegraphics[width=3.4 in,height=2.5 in]{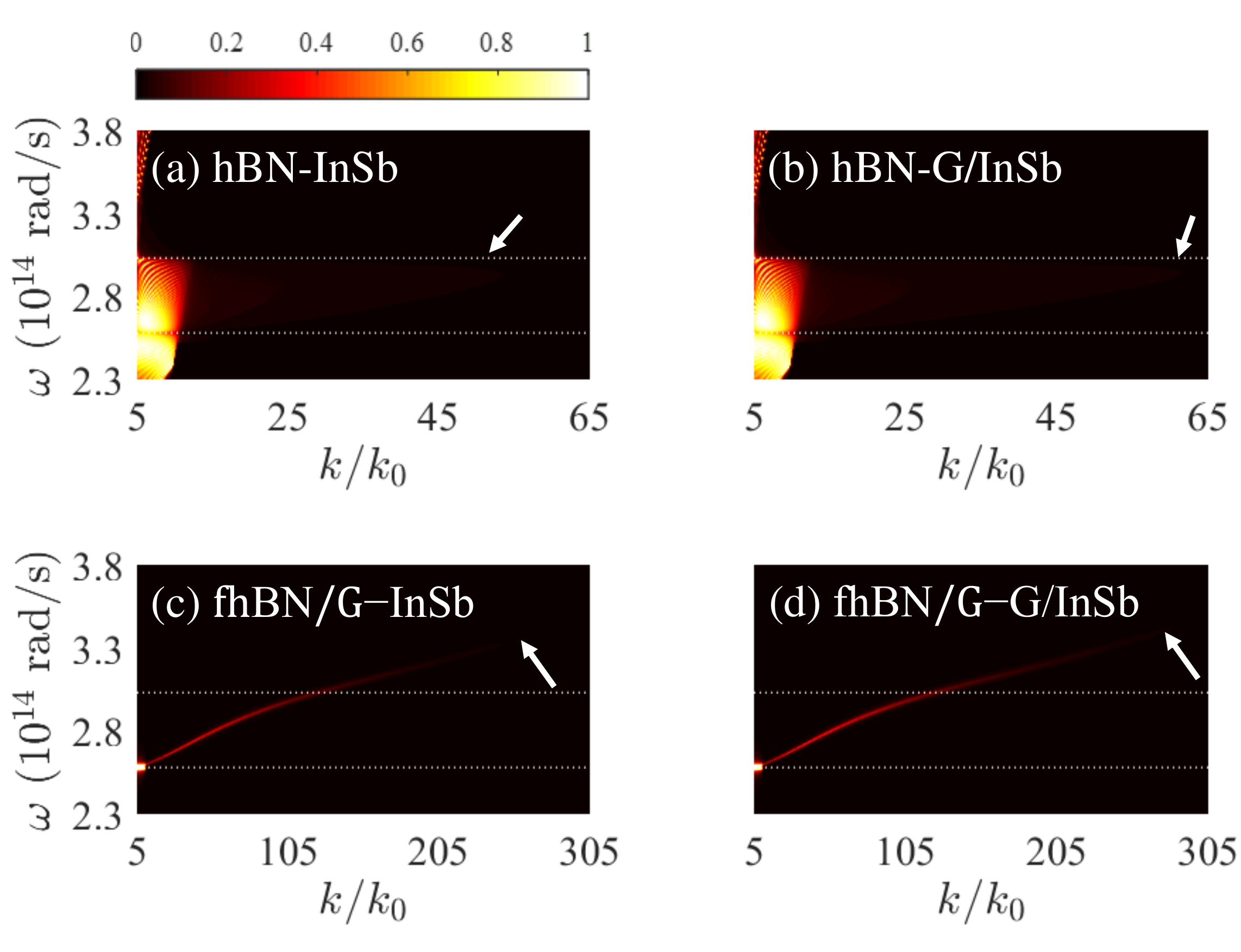}
\caption{(Color online)\ Photon tunneling probabilities $\zeta\left(\omega,k\right)$ for (a) hBN-InSb cell, (b) hBN-G/InSb cell, (c) fhBN/G-InSb cell and (d) fhBN/G-G/InSb cell. The parameters are same with those in Fig.~2. Note that the horizontal axis ranges are different for these figures. }\label{fig:photon tunneling}
\end{figure}

We further elaborate on the tunnel coupling between the emitter and the absorber by studying the photon tunneling probabilities
$\zeta\left(\omega,k\right)$ (given by Eq.~\eqref{photon tunneling}) for the four configurations as shown in Fig.~\ref{fig:photon tunneling}
for photons with frequency higher than the InSb band gap frequency, $\omega_{\rm gap}$. The bright bands shown in Fig.~\ref{fig:photon tunneling}
represent near $100\%$ photon tunneling probability, which originates from the strong coupling modes due to
various surface polaritons. The angular frequency and wavevector of these strong coupling modes satisfy the efficient tunneling condition, i.e.,
the denominator of Eq.~\eqref{photon tunneling} is minimized. It is seen from Figs.~\ref{fig:photon tunneling}(a) and \ref{fig:photon tunneling}(b)
that with the bulk h-BN being the emitter, the high photon tunneling band is mainly provided by the Reststrahlen band of type-II hyperbolic
phonon polaritons (the region between the two white dotted lines) with low wavevector. After integration over the wavevector $k$, this narrow
light band can only contribute to a small photo-induced current and limited output power. Adding a monolayer of graphene onto the InSb cell
does not considerably improve the photo-induced current and the TPV performance. In contrast, when the h-BN/graphene heterostructure film
serves as the emitter, the high photon tunneling band is no longer limited to the small wavevector region in the Reststrahlen band of type-II
hyperbolic phonon polaritons. High photon tunneling extends to very large wavevector (the deep-subwavelength evanescent wave regime)
and goes beyond the Reststrahlen band, as shown in Figs.~\ref{fig:photon tunneling}(c) and \ref{fig:photon tunneling}(d), which leads to
significantly increased phase space for high photon tunneling and hence considerably improved output power.

\begin{figure}
\centering\includegraphics[width=3.3 in,height=1.5 in]{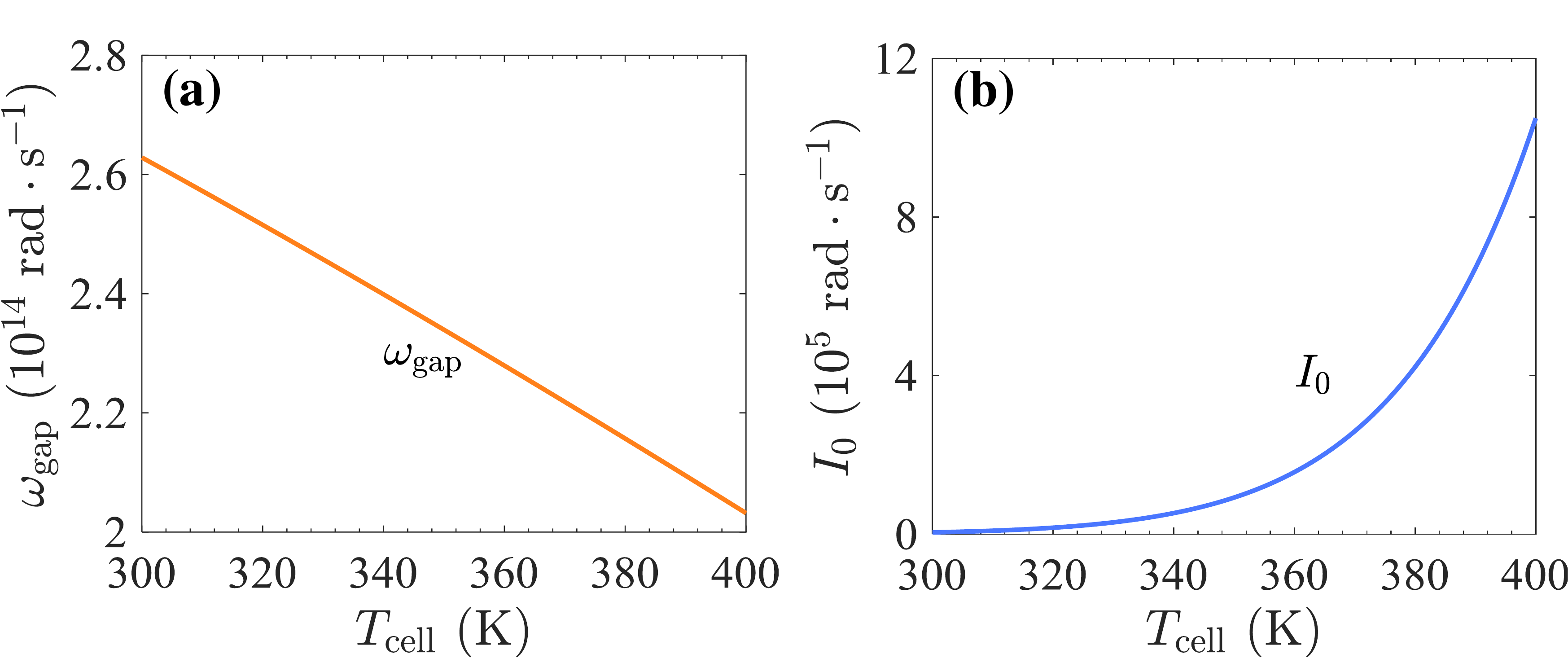}
\centering\includegraphics[width=3.3 in,height=1.5 in]{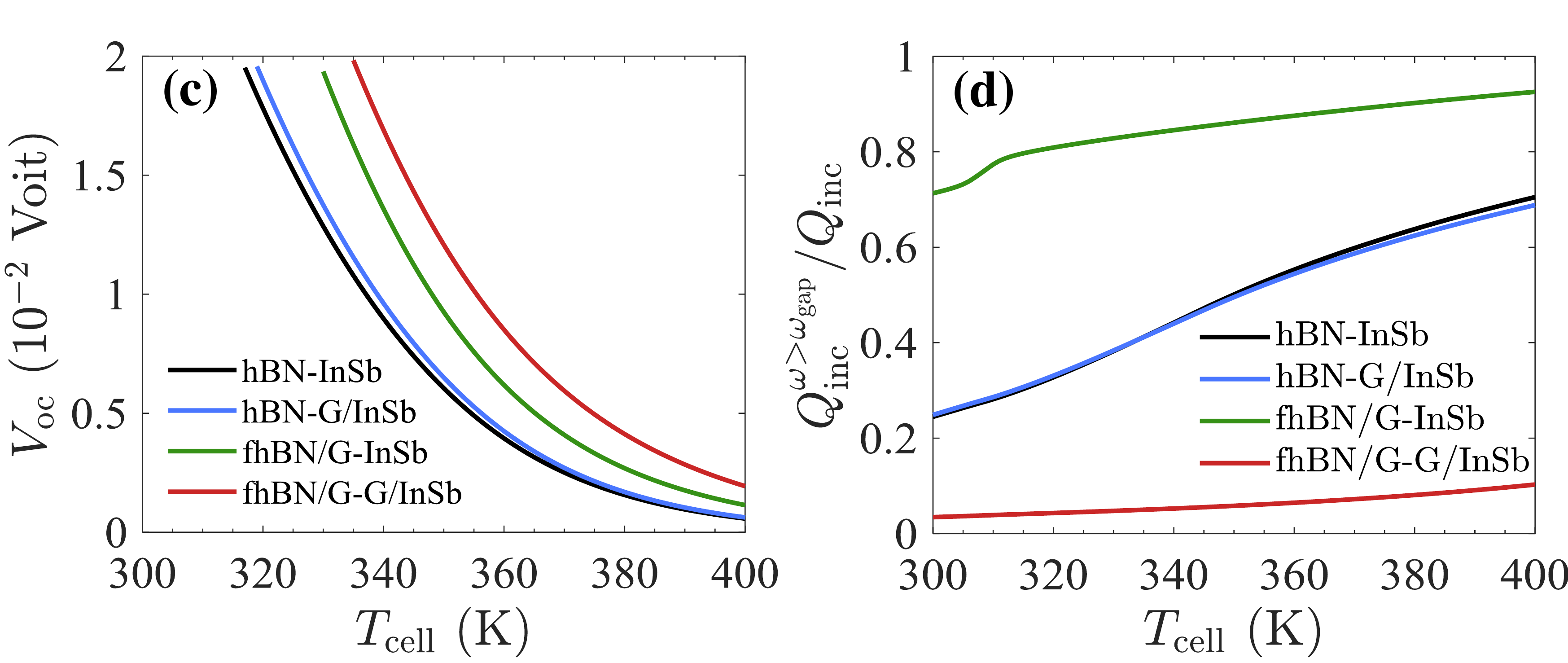}
\caption{(Color online)\ (a) Band gap frequency $\omega_{\rm gap}$ of the InSb cell, (b) reverse saturation current density $I_0$, (c) open-circuit voltage $V_{\rm oc}$ and (d) absorption fractions of incident radiation $P_{\rm inc}^{\omega>\omega_{\rm gap}}/P_{\rm inc}$ as functions of the cell temperature $T_{\rm cell}$. The temperature of the emitter is kept at $T_{\rm emit}=450$ K. The vacuum gap is $d=22$ nm, the thickness of h-BN is set as $h_{\rm bulk}=10000$ nm for bulk one, and $h_{\rm film}=20$ nm for h-BN thin film. The chemical potential of graphene is $\mu_{\rm g}=0.37$ eV.}\label{fig:effect_Tcell}
\end{figure}

To understand the performances of the four NTPV configurations under different working temperatures, we study the band gap frequency,
the reverse saturation current, the open-circuit voltage, and the absorption fraction of the incident thermal radiation as functions of the
InSb junction temperature $T_{\rm cell}$. As presented in Fig.~\ref{fig:effect_Tcell}(a), the InSb band gap frequency $\omega_{\rm gap}$
(orange solid curve) decreases from $2.6\times10^{14} \rm\ Hz$ to $2.0\times10^{14} \rm\ Hz$ when $T_{\rm cell}$ increases from
300~K to 400~K. Naively, one would expect that smaller band gap gives larger photon absorption flux and leads to better performances.
Indeed, the photo-induced current increases considerably. However, the reverse saturation current $I_0$ increases exponentially with the
cell temperature $T_{\rm cell}$ because of the reduction of the InSb band gap. As a consequence, the open-circuit voltages ($V_{\rm oc}$)
are subsequently decreased, since $V_{\rm oc}$ is given by $V_{\rm oc}=V_{\rm cell}\log\left(I_{\rm ph}/I_0+1\right)$. The total outcome
does not necessarily yield better performance.

Another important factor that dominates the energy efficiency is the absorption fraction, which is defined as the ratio of the radiative
heat flux carried by photons with frequency higher than $\omega_{\rm gap}$ to the total absorbed heat flux, i.e.,
$P_{\rm inc}^{\omega>\omega_{\rm gap}}/P_{\rm inc}$. The absorption fraction measures how much fraction of the absorbed heat flux
is useful. It is seen from Fig.~\ref{fig:effect_Tcell}(d) that the absorption fractions for the four configurations are all increased with the temperature $T_{\rm cell}$. Higher absorption fraction often leads to higher energy efficiency. It is reasonable to set the InSb cell temperature to 320~K which balances
many different aspects. The corresponding band gap is 0.17~eV and the gap frequency is $2.5\times10^{14} \rm\ Hz$.
At such a working temperature, the reverse saturation current density remains low, $I_0\simeq 1.6\times10^{4}~\rm A$.
Meanwhile, one can make good use of the incident photons, with the absorption fractions 0.80, 0.32, 0.31 and 0.043 for the hBN-InSb cell,
hBN-G/InSb cell, fhBN/G-InSb cell, and fhBN/G-G/InSb cell, respectively. Under these circumstances, the TPV cells can make the most of
the incident radiative heat and realize good performances, as shown in Fig.~\ref{fig:performance}.

The best NTPV system should have resonant photon absorption right above the band gap~\cite{Jiang2018Near},
which needs a fine matching between the emitter and the absorber. To the best of our knowledge, the fhBN/G-InSb
configuration has the best thermophotovoltaic performances due to such matching.

\begin{figure}
 \centering \includegraphics[width=3.3 in,height=1.5 in]{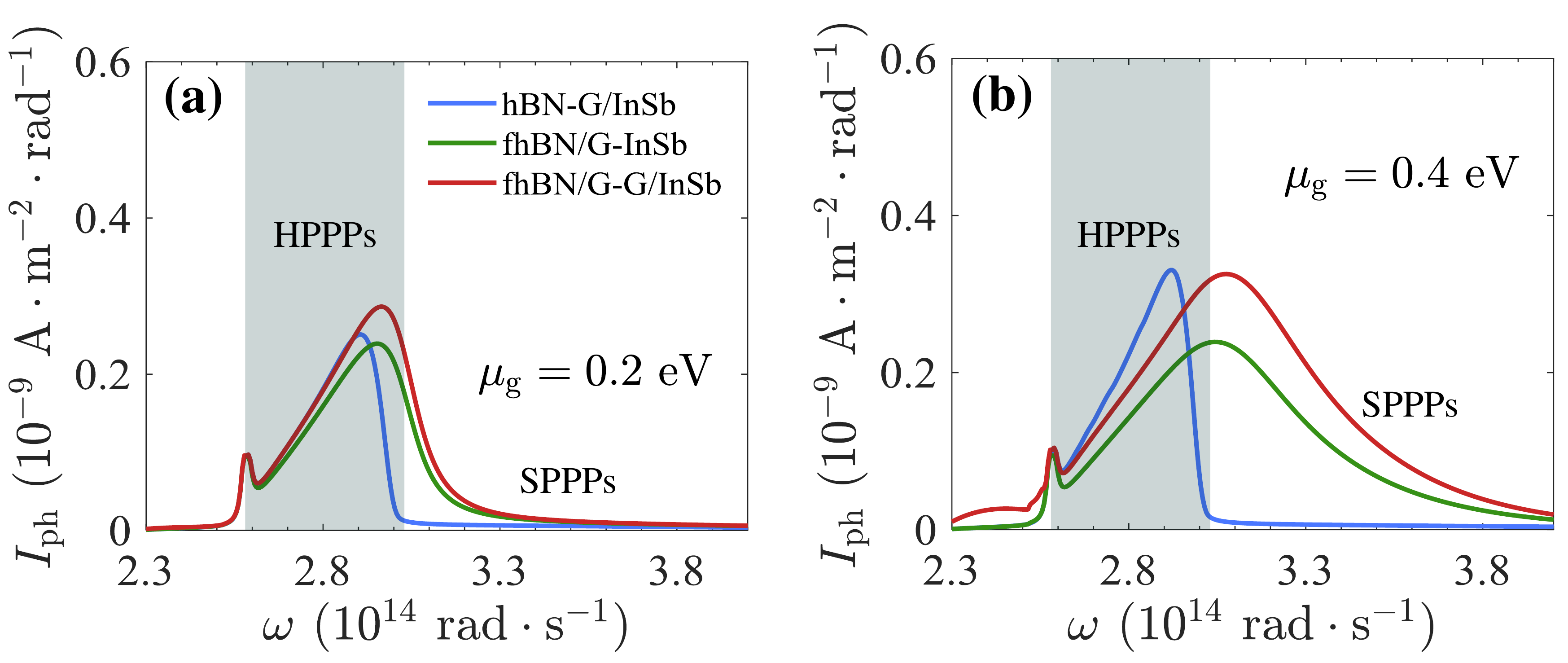}
 \centering \includegraphics[width=3.3 in,height=1.5 in]{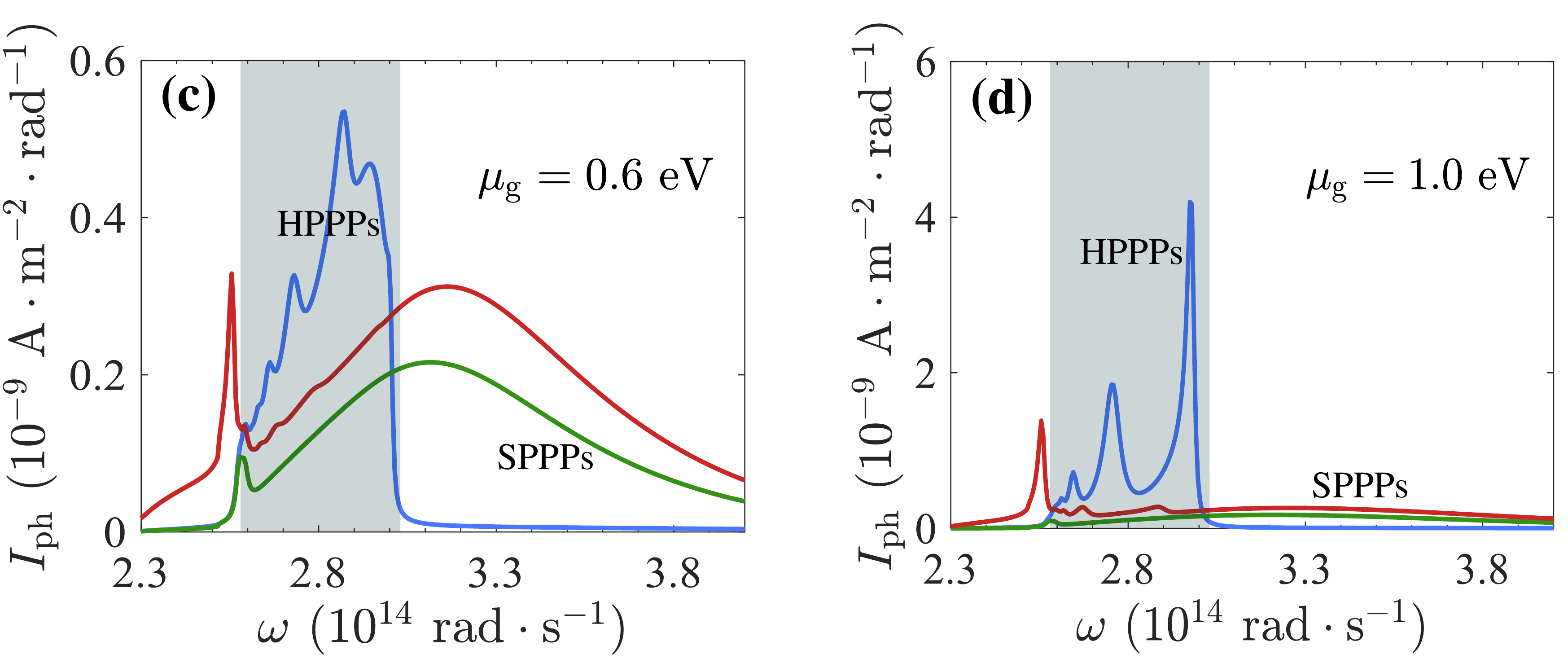}
\caption{(Color online)\ Effect of the chemical potential of graphene on the photo-induced current spectrums for hBN-G/InSb cell, fhBN/G-InSb cell and
fhBN/G-G/InSb cell at (a) $\mu_{\rm g}=0.2$ eV, (b) $\mu_{\rm g}=0.4$ eV, (c) $\mu_{\rm g}=0.6$ eV and (d) $\mu_{\rm g}=1.0$ eV. The temperatures of the emitter and the cell are set at $T_{\rm emit}=450$ K and $T_{\rm cell}=320$ K, respectively. The vacuum gap is $d=22$ nm, the thickness of h-BN is set as $h_{\rm bulk}=10000$ nm for bulk one, and $h_{\rm film}=20$ nm for h-BN thin film.}\label{fig:spectrum mug}
\end{figure}

The near-field heat transfer can be strongly modified by the chemical potential of the graphene layer, $\mu_{\rm g}$. In Fig.~\ref{fig:spectrum mug},
we plot the photo-induced current spectrums of the four different NTPV configurations with different $\mu_{\rm g}$. For the hBN-G/InSb cell,
with large chemical potentials, the resonant peak in the HPPPs region of the photo-induced current
spectrum is enhanced and split into small side peaks due to the resonant coupling with graphene plasmons. For the large chemical potential
$\mu_{\rm g}=1.0$~eV, the near field coupling between the SPhPs and the graphene plasmons is significantly enhanced.
These behaviors are due to the dependence of the graphene plasmon frequencies on
the chemical potential, as demonstrated in Refs.~\onlinecite{Dai_hBN}, \onlinecite{bnnt2} and
\onlinecite{messina2013graphene}. The enhancement of photo-induced current spectrum by tuning $\mu_{\rm g}$
provides an effective way to further enhance the performances of the graphene-based NTPV systems.

\subsection{Optimization}\label{optimization}
\begin{figure}
 \centering \includegraphics[width=3.3 in,height=1.60 in]{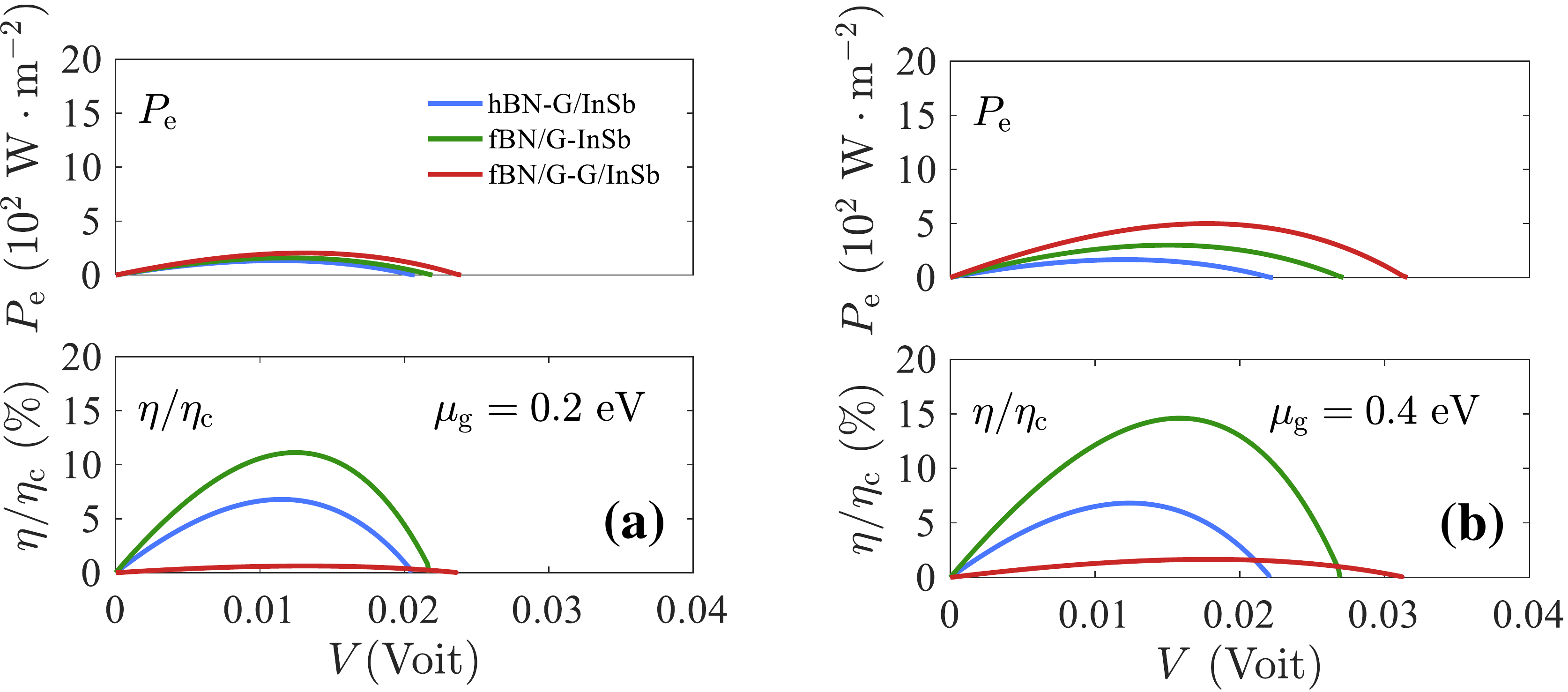}
 \centering \includegraphics[width=3.3 in,height=1.60 in]{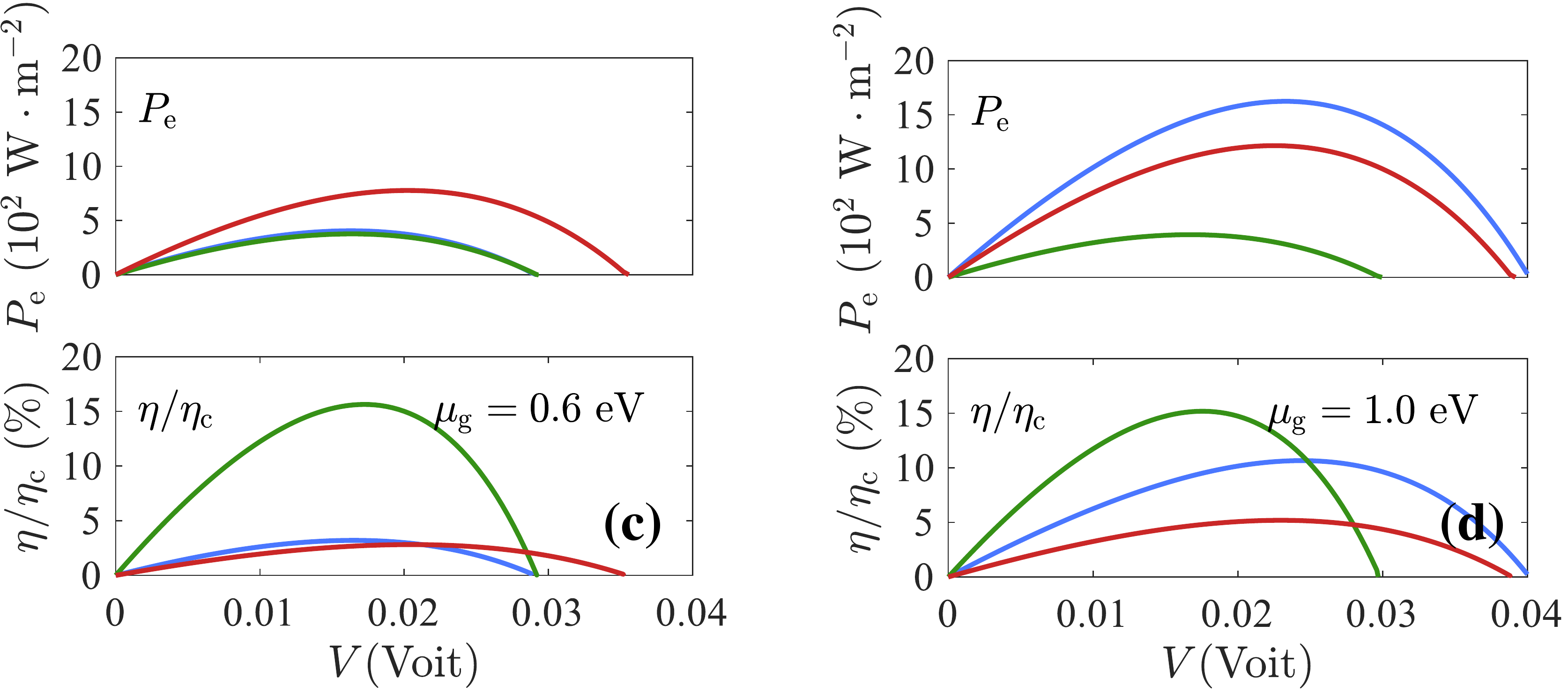}
\caption{(Color online)\ Effect of the chemical potential of graphene on the performances for hBN-G/InSb cell, fhBN/G-InSb cell and fhBN/G-G/InSb cell at (a) $\mu_{\rm g}=0.2$ eV, (b) $\mu_{\rm g}=0.4$ eV, (c) $\mu_{\rm g}=0.6$ eV and (d) $\mu_{\rm g}=1.0$ eV. Other parameters are same with those in Fig.~\ref{fig:spectrum mug}.}\label{fig:performance mug}
\end{figure}

\begin{figure}
\centering\includegraphics[width=3.3 in,height=1.6 in]{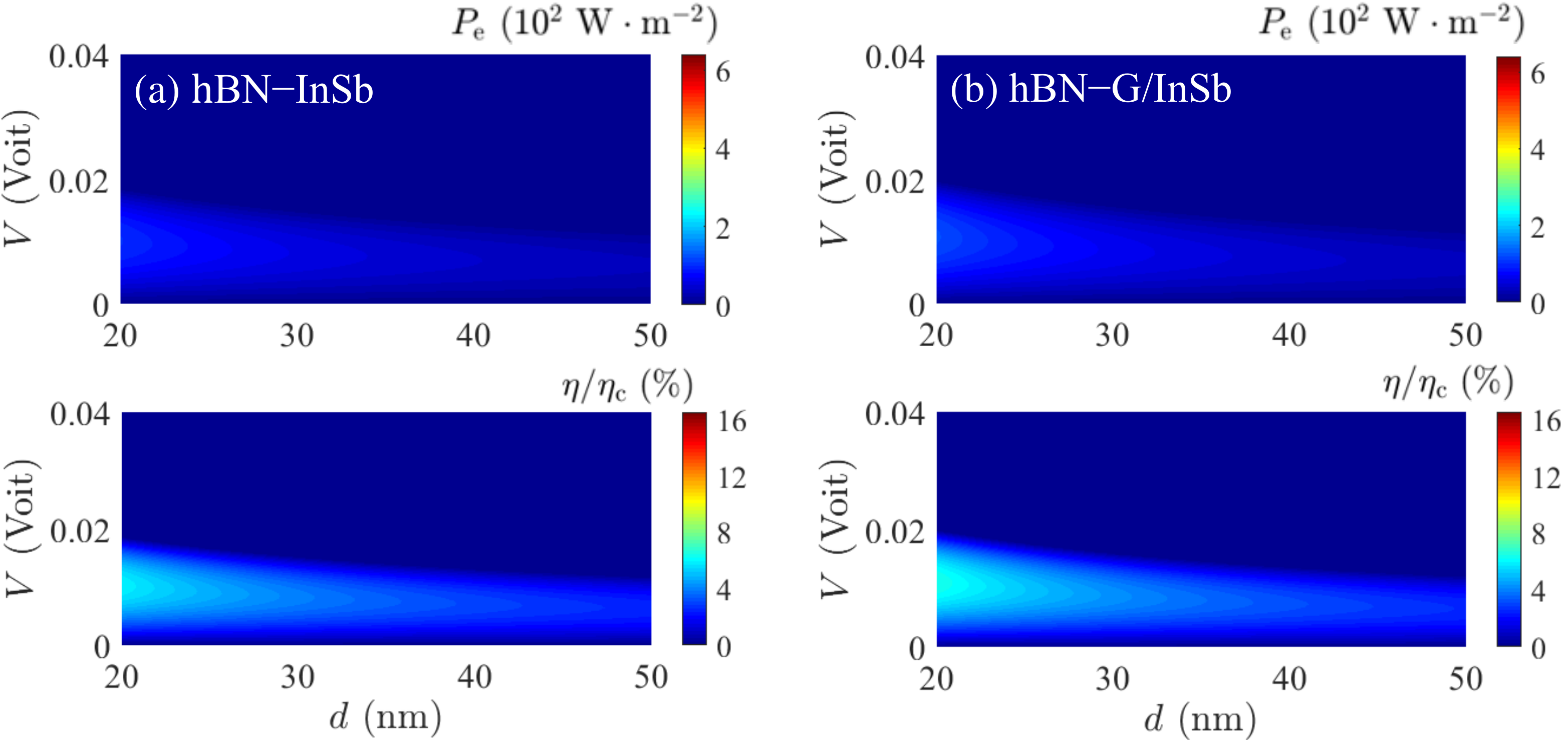}
\centering\includegraphics[width=3.3 in,height=1.6 in]{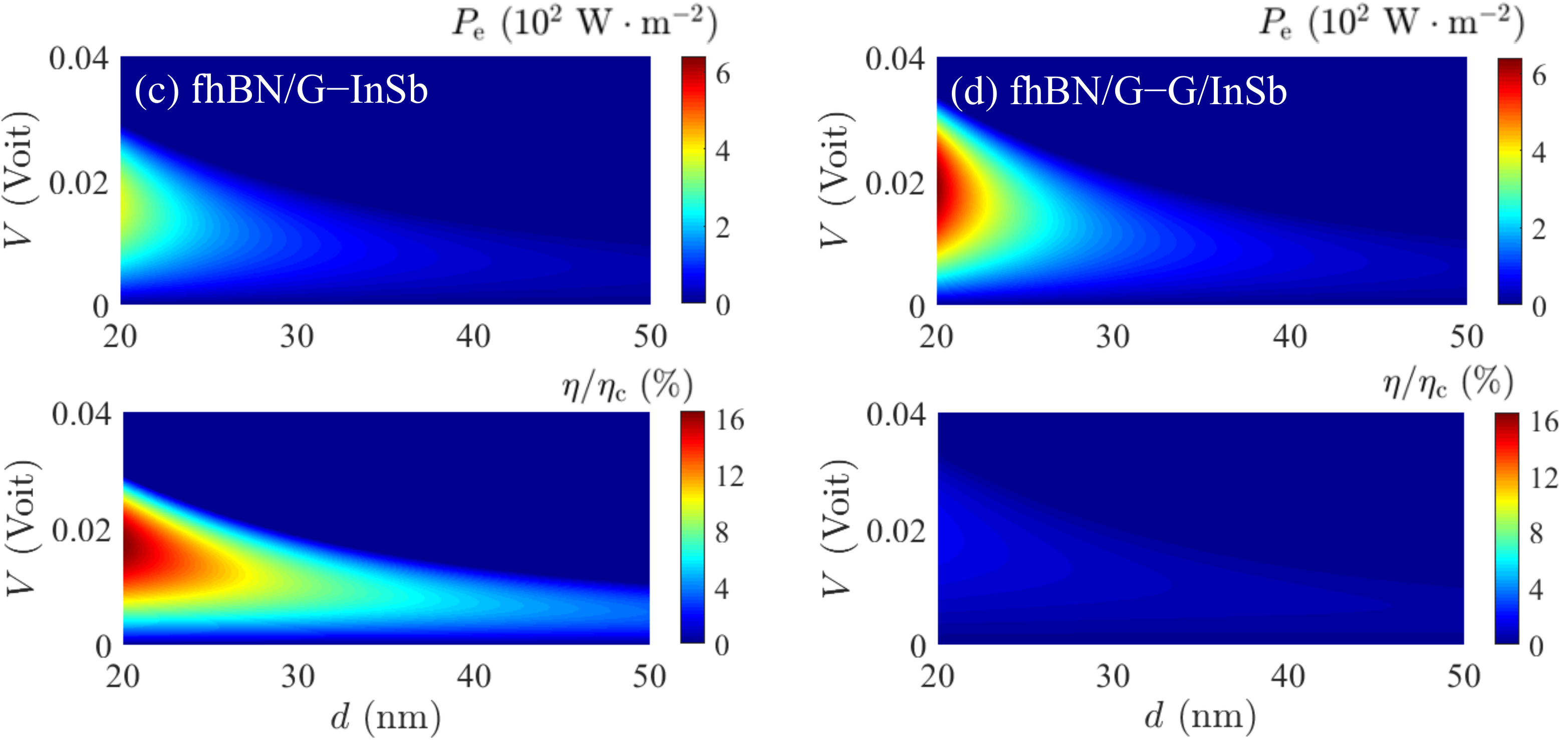}
\caption{(Color online)\ Effect of the vacuum gap $d$ on the performances for (a) hBN-InSb cell, (b) hBN-G/InSb cell, (c) fhBN/G-InSb cell and (d) fhBN/G-G/InSb cell. The temperatures of the emitter and the cell are set at $T_{\rm emit}=450$ K and $T_{\rm cell}=320$ K. The thickness of h-BN is set as $h_{\rm bulk}=10000$ nm for bulk one, and $h_{\rm film}=20$ nm for h-BN thin film. The chemical potential of graphene is $\mu_{\rm g}=0.37$ eV.}\label{fig:effcet d}
\end{figure}

\begin{figure}
\centering\includegraphics[width=3.3 in,height=1.6 in]{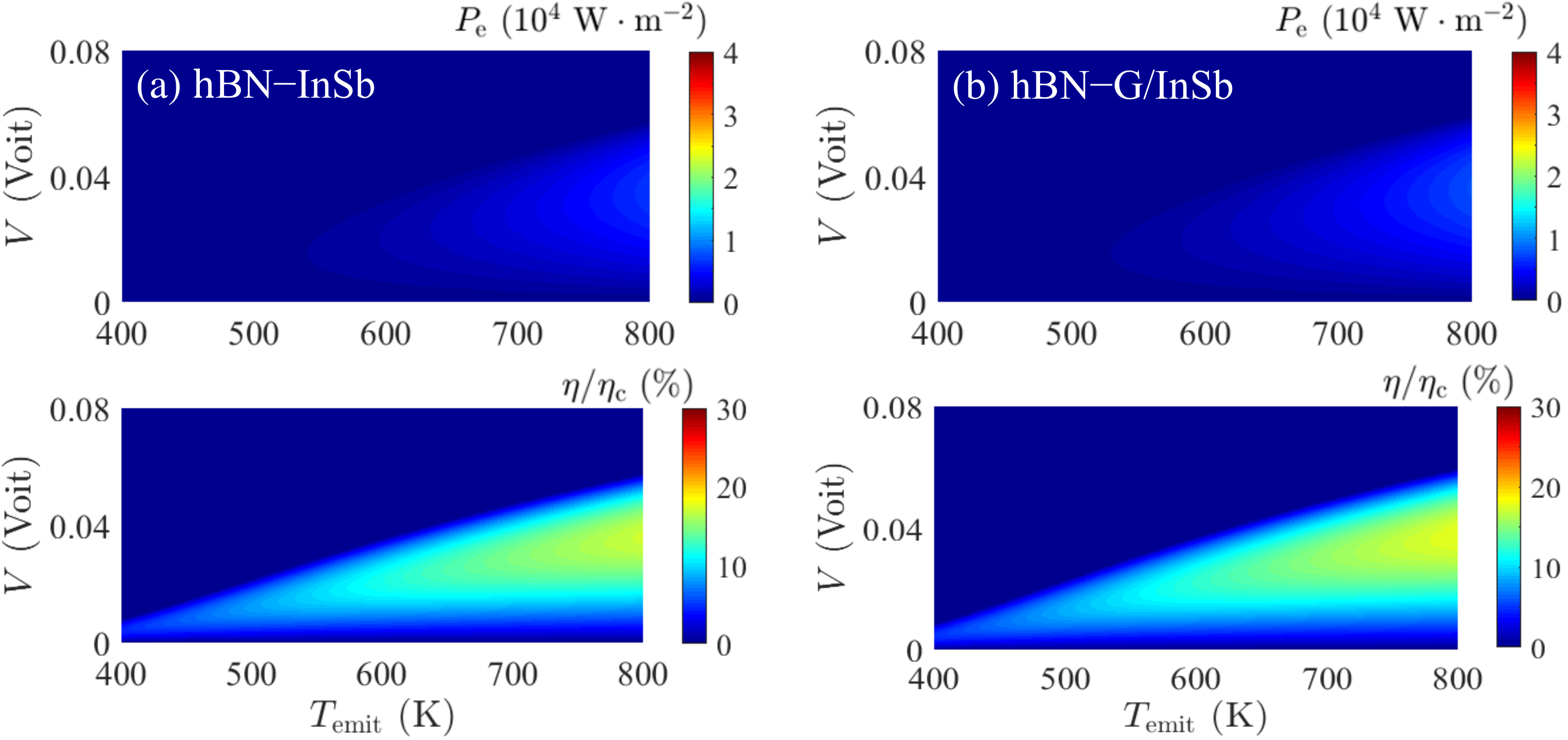}
\centering\includegraphics[width=3.3 in,height=1.6 in]{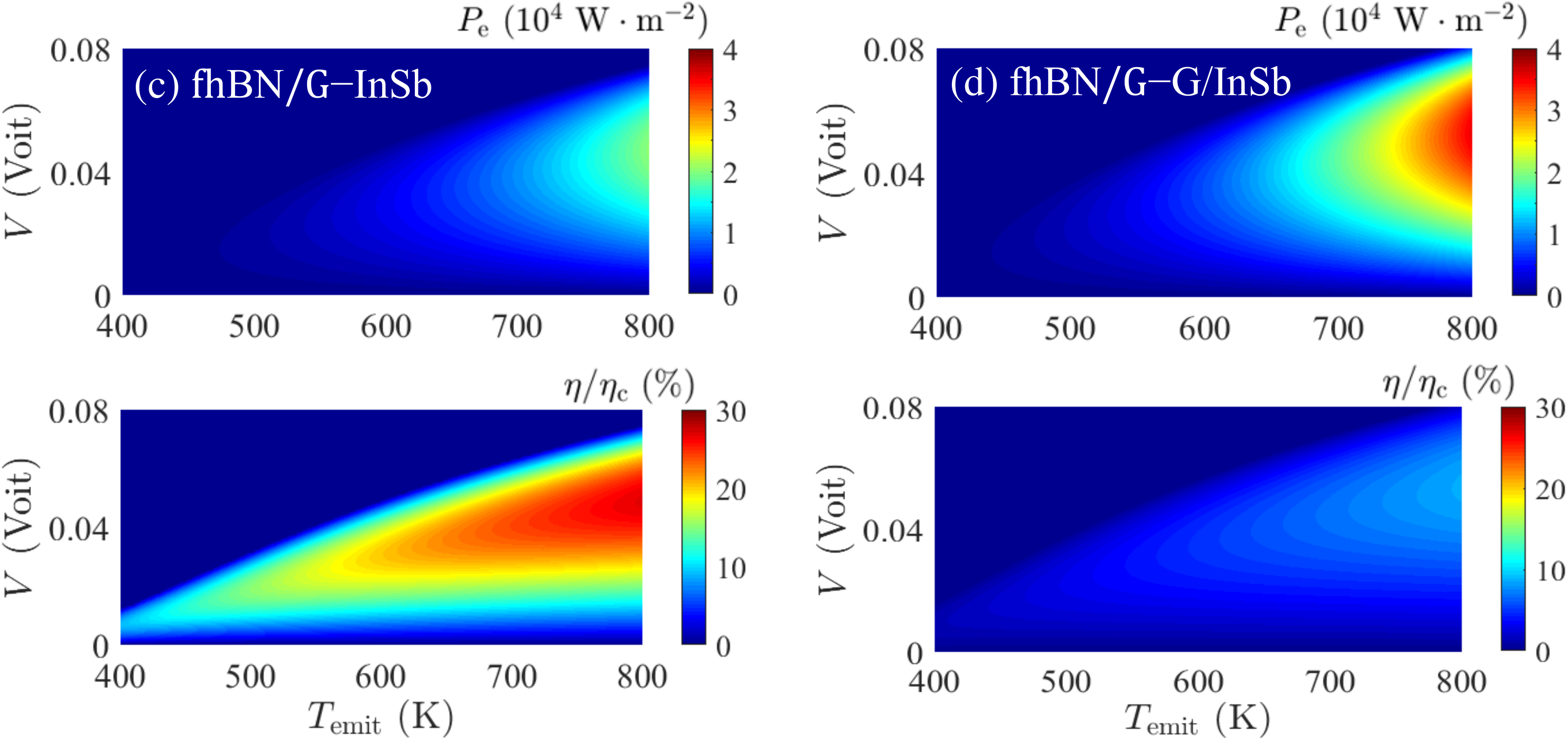}
\caption{(Color online)\ Effect of the emitter temperature $T_{\rm emit}$ on the performances for (a) hBN-InSb cell, (b) hBN-G/InSb cell (c) fhBN/G-InSb cell and (d) fhBN/G-G/InSb cell. The temperature of the cell is kept at $T_{\rm cell}=320$ K. The thickness of h-BN is set as $h_{\rm bulk}=10000$ nm for bulk one, and $h_{\rm film}=20$ nm for h-BN thin film. The vacuum gap is $d=20$ nm, and the chemical potential of graphene is $\mu_{\rm g}=0.37$ eV.}\label{fig:effcet Temit}
\end{figure}

The output power and energy efficiency for the three configurations with graphene at four different chemical potentials are shown
in Fig.~\ref{fig:performance mug}. The output powers and energy efficiencies for these three configurations are all improved when the
chemical potential of graphene is increased. Especially for the fhBN/G-G/InSb cell, the maximum output power density at $\mu_{\rm g}=1.0$ eV is
increased to nearly 6 times of that at $\mu_{\rm g}=0.2$~eV. Meanwhile, the corresponding maximum efficiency is enhanced to 8.3 times of that
at $\mu_{\rm g}=0.2$~eV. For the hBN-G/InSb cell, the maximum output power density at $\mu_{\rm g}=1.0$ eV is
enhanced to about 12 times of that at $\mu_{\rm g}=0.2$~eV, while the the corresponding maximum efficiency is enhanced to 1.6 times of that
at $\mu_{\rm g}=0.2$~eV. Also for the fhBN/G-InSb cell, the maximum output power density and maximum efficiency at higher chemical potential are
enhanced to 2.5 and 1.4 times of those at $\mu_{\rm g}=0.2$~eV, respectively. Such significant improvements of the TPV
performances of the fhBN/G-InSb, hBN-G/InSb and fhBN/G-G/InSb cells are in accordance with the strongly enhanced photo-induced current spectrums shown in Fig.~\ref{fig:spectrum mug}.

We also study the influence of the vacuum gap $d$ and the temperature of the emitter $T_{\rm emit}$ on the performances of four configurations,
as shown in Figs.~\ref{fig:effcet d} and \ref{fig:effcet Temit}, respectively. The electric power density and the normalized energy efficiency are
remarkably enhanced when the vacuum gap is reduced to 20 nm or when the emitter temperature is increased to 800 K. These results
demonstrate that the near-field heat transfer is crucial for the enhancement of both the energy efficiency and the output power.

The optimal output power is achieved in the fhBN/G-G/InSb cell, with a power density of $3.5\times10^{4} \rm\ W/\rm m^2$.
The optimal efficiency, $27\%$ of the Carnot efficiency, is realized in the fhBN/G-InSb cell, while a moderate output power density
$2.0\times10^{4} \rm\ W/\rm m^2$ is realized as well. Therefore, the best
balance between the energy efficiency and the output power comes from the configuration with fhBN/G-InSb cell. The overall
trend is that narrow vacuum gap, proper chemical potential of graphene, and high temperature of the emitter are favorable
for high TPV performances.

\section{Conclusions}\label{conlusions}

We investigate the energy efficiency and output power of four different NTPV systems, denoted as the hBN-InSb cell, hBN-G/InSb cell, fhBN/G-InSb
cell, and the fhBN/G-G/InSb cell, where the SPhPs in h-BN and graphene plasmons as well as their coupling play viable roles in enhancing
the energy efficiency and power of these NTPV systems. It is found that the optimal output electric power with $3.5\times10^{4} \rm\ W/\rm m^2$ is
achieved in hBN/graphene-graphene/InSb cell. While the optimal efficiency, $27\%$ of the Carnot efficiency, is reached in h-BN/graphene-InSb cell.
The performance of the h-BN/graphene based cell can be further improved by tuning the chemical potential of graphene. Combining with the fact of
the experimental availability of the h-BN/graphene heterostructure and the state-of-art doping of graphene~\cite{lu2017synthesis,zhang2018direct},
h-BN/graphene based cell can be useful for future thermophotovoltaic systems with high performances.

Remarkably, using such a graphene-BN-InSb near-field heterostructure design, we show that the performance of
our NTPV systems can be comparable with the state-of-art thermoelectric systems working in the
same temperature range. For instance, the state-of-art output power density
of thermoelectric generators is realized in Ref.[~\onlinecite{highPF}] with a power factor (PF) $PF=2.5\times 10^5$~W/m$^2$ for a device
working between two baths with temperatures $T_{\rm h}=873$~K and $T_{\rm c}=330$~K, whereas for the same conditions
our NTPV device gives $PF=4.0\times 10^4$~W/m$^2$. The state-of-art device efficiency of thermoelectric generators is
given in Ref.[~\onlinecite{zhao2015science}] where $\eta=27\%\eta_{\rm c}$ for a thermoelectric generator working
between two thermal reservoirs with temperatures $T_{\rm h}=773$~K and $T_{\rm c}=300$~K. The corresponding device
$ZT$ is $Z_{\rm dev}T=1.34$. In comparison, for the same temperatures, our
NTPV device can also give $\eta=27\%\eta_{\rm c}$. These comparisons indicate that the NTPV systems have
a high potential for future thermal to electric energy conversion.

\section{Acknowledgments}
R.W., J.L., and J.-H.J. acknowledge support from the National Natural Science Foundation of China (NSFC Grant No. 11675116), the Jiangsu distinguished professor funding and a Project Funded by the Priority Academic Program Development of Jiangsu Higher Education Institutions (PAPD). R.W. thanks Professor Chen Wang for discussions.

\bibliography{ghNTPV_v1}
\end{document}